\documentclass[%
superscriptaddress,
 amsmath,amssymb,
 aps,
 jcp,
 letterpaper,
]{revtex4-2}

\usepackage{tikz}
\usetikzlibrary{trees}
\usetikzlibrary{decorations.pathmorphing}
\usetikzlibrary{decorations.markings}
\tikzset{
    photon/.style={decorate, decoration={snake,segment length=1.5mm}, draw=black},
    electron/.style={draw=black, postaction={decorate},
        decoration={markings,mark=at position .55 with {\arrow[draw=black]{>}}}}, 
    gluon/.style={decorate, draw=magenta,
        decoration={coil,amplitude=4pt, segment length=5pt}},
    boundelectron/.style={thick, double}
}

\usepackage{graphicx}
\usepackage{dcolumn}
\usepackage{bm}

\newcolumntype{.}{D{.}{.}{8}}

\usepackage[utf8]{inputenc}
\usepackage{physics}    
\usepackage{amsmath, amssymb}
\usepackage{tabularx,booktabs,array,dcolumn}
\usepackage{setspace}
\usepackage{vmargin}
\usepackage{xcolor}
\usepackage{graphicx,psfrag,subfigure}

\usepackage{float}
\usepackage[all]{xy}

\usepackage{bm}
\usepackage{mathtools}
\usepackage{physics}
\usepackage[english]{babel}

\def\som{Supplementary Material}

\usepackage{bm}
\usepackage{mathtools}
\usepackage{physics}

\newcommand{\bos}[1]{\boldsymbol{#1}}

\newcommand{\cm}{cm$^{-1}$}

\newcommand{\Tt}{^\mathrm{T}}

\usepackage[unicode]{hyperref}
\hypersetup{
   unicode=true,          
   plainpages=false,
   colorlinks=true,       
   citecolor=blue,        
}

\newcommand{\pd}[2]{{\frac{\partial #1}{\partial #2}}}

\def\cpl{C}
\def\Sgp{$^1\Sigma_\text{g}^+$}

\def\epsi{\varepsilon}
\def\hel{H_\text{el}}
\def\ordo{\mathcal{O}}
\def\vtheta{\vartheta}
\def\Eh{$\mathrm{E}_\mathrm{h}$}

\def\XX{\emph{X}}
\def\EF{\emph{EF}}
\def\GK{\emph{GK}}
\def\HH{\text{\emph{H$\bar{H}$}}}
\def\Sfive{\emph{S5}}
\def\Ssix{\emph{S6}}
\def\pP{\mathcal{P}}
\def\kk{{(k)}}
\def\minC{k_0}
\def\maxC{k_C}

\usepackage{soul}

\begin{document}

\title{%
Vibronic mass computation for the $EF$--$GK$--$H\bar{H}$ $^1\Sigma_\text{g}^+$ manifold of molecular hydrogen
}

\author{Edit M\'atyus} 
\email{edit.matyus@ttk.elte.hu}
\author{D\'avid Ferenc} 
\affiliation{ELTE, Eötvös Loránd University, Institute of Chemistry, 
Pázmány Péter sétány 1/A, Budapest, H-1117, Hungary}

\date{\today}

\begin{abstract}
\noindent
A variational procedure is described for the computation of the non-adiabatic mass correction tensor applicable for multi-dimensional electronic manifolds. 
The 30-year old computations of Wolniewicz, Dressler and their co-workers are appended with the computed vibronic mass functions corresponding to the \EF--\GK--\HH--\Sfive--\Ssix\ \Sgp\ manifold of the hydrogen molecule. 
Initial results are reported for the vibronic energies including effective vibronic masses. 
Necessary further improvements and further developments are discussed.


\end{abstract}
\maketitle

\clearpage
\section{Introduction}
This paper is dedicated to the memory of L. Wolniewicz. \\

Wolniewicz, Dressler and their co-workers published a series of papers on the theoretical (and experimental) study of the electronic-vibrational-rotational spectroscopy of molecular hydrogen  \cite{WoDr77,DrGaQuWo79,SeQuDr88,DrWo86,QuDrWo90,WoDr92,WoDr94,YuDr94}. In their work, the rovibronic wave function was expanded in terms of sum of products of electronic and rovibrational basis functions.

This approach can, in principle, converge to the exact non-relativistic result, if a sufficiently large electronic and vibrational basis set is included in the computation. In a practical computation, the basis set always has to be truncated. 
Yu and Dressler \cite{YuDr94}  included five coupled $^1\Sigma_\text{g}^+$ states (labelled as \EF, \GK, \HH, \Sfive, and \Ssix) \cite{WoDr94} to describe the vibronic ($J=0$) states. For the rovibronic ($J>0$) computations, the electronic space was extended with $^1\Pi$ and $^1\Delta$ wave functions.

It has been pointed out, \emph{e.g.,} Refs.~\cite{Te03,PaSpTe07,MaTe19} that the electronic basis truncation error in the approximate (ro)vibronic energies is of order $\ordo(\epsi)$ that may be too large for spectroscopic purposes, where $\epsi^2=m_\text{el}/m_\text{p}$
is the electron-to-proton mass ratio. A better approximation is obtained if the effect of the discarded electronic subspace is perturbatively corrected. 
The effective rovibronic Hamiltonian that is an order of magnitude better, \emph{i.e.,} $\ordo(\epsi^2)$, contains a kinetic or as it is also called `mass-correction' term.
The effective Hamiltonian for a multi-dimensional electronic manifold has been recently formulated \cite{MaTe19}
following Refs.~\cite{Te03,PaSpTe07}. 

The mass-correction term appearing in the $\ordo(\epsi^2)$ effective Hamiltonian contains the reduced resolvent of the electronic Hamiltonian, for which a direct summation over the electronic basis may return inaccurate results.
In the present work, we adopt Schwartz' method \cite{Sc61}, commonly used for the precise evaluation of the sum-over-states expression in the Bethe logarithm, for the computation of the mass-correction matrix. The sum-over-states contribution, is represented by a `perturbed' wave function that can be systematically improved through the minimization of an appropriate functional. 

The vibronic mass correction is evaluated for single and multi-dimensional electronic subspaces corresponding to the \EF, \GK, \HH, \Sfive, and \Ssix\ \Sgp\ electronic manifold of molecular hydrogen. The computed vibrational mass correction functions are used to complement the work of Wolniewicz, Dressler, and their co-workers, and initial results are reported for the vibronic energies corresponding to effective vibronic masses.

\section{Nuclear-motion Hamiltonian over coupled electronic states \label{ch:coupled}}
Eigenstates of the $\widehat{H}$ (non-relativistic) electron-nucleus Hamiltonian are 
approximated over an ansatz of products of electronic and vibrational functions.
In this work, we assume an adiabatic electronic basis, $\psi_1(R),\ldots,\psi_d(R)$ ($R$ labels the nuclear coordinates), \emph{i.e.,} the electronic basis functions are eigenfunctions of the electronic Hamiltonian, 
\begin{align}
  H(R) \psi_\alpha(R) 
  =
  E_\alpha(R) \psi_\alpha(R) \; ,
\end{align}
but only a finite, $d$-dimensional electronic subspace 
\begin{align}
  P=\sum_{\alpha=1}^d P_\alpha 
  \quad\text{with}\quad 
  P_\alpha = |\psi_\alpha(R) \rangle \langle \psi_\alpha(R)|
  \label{eq:Psubspace}
\end{align}
is retained for the computation, where $\langle \psi_\alpha(R)|\psi_\beta(R)\rangle=\delta_{\alpha,\beta}$. 
The Hamiltonian for the quantum nuclear motion over the $P$ electronic subspace reads as
\begin{align}
  (\widehat{\bos{H}}_P)_{\alpha\beta}
  =
  \langle 
    \psi_\alpha|\widehat{H}_P | \psi_\beta
  \rangle
  =
  \langle 
    \psi_\alpha|\widehat{K} | \psi_\beta
  \rangle
  +
  \langle 
    \psi_\alpha| H | \psi_\beta
  \rangle  \; ,
  \label{eq:cplham}
\end{align}
where the nuclear kinetic energy part is
\begin{align}
  \langle
    \psi_\alpha | \widehat{K}| \psi_\beta
  \rangle
  =
  -\sum_i
  \left(%
    \frac{1}{2}(\epsi\partial_i)^2 \delta_{\alpha\beta} 
    +\epsi \langle \psi_\alpha|\partial_i \psi_\beta \rangle (\epsi\partial_i)
    +\frac{\epsi^2}{2}
    \langle \psi_\alpha| \partial_i^2 \psi_\beta \rangle
  \right) \; .
  \label{eq:cplkeo}
\end{align}
If the $P$ electronic subspace is separated by a finite gap from the rest of the electronic spectrum 
over the nuclear coordinates relevant for the nuclear dynamics, then qualitatively correct energy estimates
can be expected from this truncated, $d$-dimensional description, Eq.~(\ref{eq:Psubspace}). 
Eigenvalues of $\widehat{\bos{H}}_P$ approximate the exact electron-nucleus non-relativistic energies
with an error of $\ordo(\epsi)$ \cite{Te03,MaTe19}. 

Better estimates can be obtained if the electronic states which are not included in the $P$
subspace are accounted for perturbatively. 

To obtain an $\epsi$-order more accurate energies, 
a $\widehat{\mathcal{H}}_P^{(2)}$ second-order effective Hamiltonian operator
has been formulated \cite{MaTe19} following earlier work \cite{Te03,PaSpTe07}.
The matrix representation of this effective Hamiltonian for the rovibrational motion, over the $\psi_\alpha$ ($\alpha=1,\ldots,d$) adiabatic basis is 
\begin{align}
  (\widehat{\bos{H}}_P^{(2)})_{\alpha\beta}
  &=
  \langle 
    \psi_\alpha|\widehat{H}_P^{(2)} | \psi_\beta
  \rangle
  \nonumber \\
  &=
  \langle 
    \psi_\alpha|\widehat{K} | \psi_\beta
  \rangle
  +
  \langle 
    \psi_\alpha| H | \psi_\beta
  \rangle 
  +
  \frac{\epsi^2}{2}
  \sum_{j,i} \sum_{a,b}
  (\epsi\partial_j)
  M_{\alpha\beta,ij} 
  (\epsi\partial_i)
  \label{eq:nadham}
\end{align}
where $\partial_{j}=\partial/\partial R_{j}$ labels numerical differentiation with respect to the $j$th nuclear coordinate and
the `so-called' mass-correction matrix is 
\begin{align}
  M_{\alpha\beta,ij}
  =
  \sum_{a,b=1}^d
  \langle 
      \psi_\alpha | 
      P_a (\partial_j P) 
      (\mathcal{R}_a+\mathcal{R}_b)  
      (\partial_i P) P_b |
      \psi_\beta
   \rangle \;, \label{eq:massmulti}
\end{align}
and $\mathcal{R}_i=(H(R)-E_i(R))^{-1}P^\perp$ is the reduced resolvent with $P^\perp=1-P$. 
If $\psi_1,\ldots,\psi_d$ form an adiabatic basis set, then
the mass-correction matrix simplifies to 
\begin{align}
  M_{\alpha\beta,ij}
  &=
  \langle
    \partial_j \psi_\alpha |
    \mathcal{R}_\alpha
    +
    \mathcal{R}_\beta |
    \partial_i \psi_\beta
  \rangle \nonumber \\
  &=
  \langle
    \partial_j \psi_\alpha |
    (H(R)-E_\alpha(R))^{-1}P^\perp |
    \partial_i \psi_\beta
  \rangle \nonumber \\      
  &\quad +
  \langle
    \partial_j \psi_\alpha |
    (H(R)-E_\beta(R))^{-1}P^\perp |
    \partial_i \psi_\beta
  \rangle \; . \label{eq:massmultiad}
\end{align}
If a single $\psi_\alpha$ electronic state spans the $P$ active space ($d=1$), the expression simplifies to
\begin{align}
  M_{\alpha\alpha,ij}
  =
  2
  \langle 
    \partial_j \psi_\alpha | 
      \mathcal{R}_1
    |\partial_i \psi_\alpha
  \rangle
  =
  2
  \langle 
    \partial_j \psi_\alpha | 
      (H-E_\alpha)^{-1}(1-P_\alpha)
    |\partial_i \psi_\alpha
  \rangle  \; . \label{eq:massone}
\end{align}
The single-state mass correction, Eq.~(\ref{eq:massone}), has been formulated several times \cite{HeAs66,HeOg98,PaKo09,prx17}
and was successfully used in spectroscopic applications \cite{Sch01H2p,Sch01H2O,PaKo09,CzPuKoPa18,KoPuCzLaPa19,FeMa19HH,FeKoMa20}. 
We are not aware of any computation with the multi-state expression, Eqs.~(\ref{eq:massmulti}) or (\ref{eq:massmultiad}).

In Sec.~\ref{ch:method}, a variational approach is described for the evaluation of $M_{\alpha\beta,ij}$.
In Sec.~\ref{ch:vibronic}, 
we compute the vibronic mass corrections for the example of the lowest electronically excited \Sgp\ 
electronic manifold of molecular hydrogen,
and report initial results for the vibronic energy by including the effective vibronic masses.
The paper ends with an outlook to further computational, algorithmic and theoretical work.

\section{Evaluation of the mass correction matrix using Schwartz' method \label{ch:method}}
We adapt Schwartz' method, originally proposed to compute the sum-over-states expression in the non-relativistic Bethe logarithm \cite{Sc61}, for the evaluation of the mass matrix elements.
The method is first adopted for the ground electronic state. 
Then, orthogonality constraints and lower boundedness of the functional is discussed  for an electronically excited state.  
In the last step, the evaluation of the mass correction for a multi-dimensional electronic subspace is explained.

\subsection{Non-adiabatic mass for the ground electronic state \label{ch:gs}}
A non-adiabatic mass matrix element for the ground electronic state ($E_0,\psi_0$) reads as
\begin{align}
  M_{00,ij}
  &=
  \langle
    \partial_{i} \psi_0 | (H-E_0)^{-1} P_0^\perp | \partial_{j} \psi_0 
  \rangle 
  \nonumber \\
  &=
  \sum_{n\neq 0}
  \frac{%
  \langle
    \partial_{i} \psi_0 | \psi_n \rangle \langle \psi_n | \partial_{j} \psi_0 
  \rangle
  }{E_n-E_0}  \; ,
  \label{eq:sos}
\end{align}
where the summation includes also integration. 
Following Schwartz, we re-write the sum-over-states expression, as
\begin{align}
  M_{00;ij}
  =
  \langle %
    \partial_{i}\psi_0 | \phi^{(0,j)}_0
  \rangle \; ,
  \label{eq:rewrite}
\end{align}
where the $\phi^{(0,j)}_0$ `perturbed' wave function minimizes the functional 
(henceforth, the 0 and $(0,j)$ sub- and superscripts are suppressed for brevity):
\begin{align}
  \mathcal{W}[\phi,\lambda]
  =
  \langle \phi |(H-E_0)| \phi \rangle 
  - 
  2\langle \phi | \partial \psi_0 \rangle
  -
  \lambda \langle \phi | P_0 | \phi \rangle \;,
  \label{eq:Hyfun}  
\end{align}
and $\lambda$ is a Lagrange multiplier introduced to ensure that 
$\phi$ is orthogonal to $\psi_0$, and
$P_0=|\psi_0\rangle\langle \psi_0|$.
In general, the $\langle\phi|\psi_0\rangle=0$ orthogonality constraint is not automatically fulfilled, 
unless some special symmetry condition applies, and it is necessary
to ensure the $n\neq0$ condition in the sum-over-states expression, Eq.~(\ref{eq:sos}).

We need to find the stationary point (minimum) of $\mathcal{W}$ with respect to the variation 
of $\phi$ and $\lambda$:
\begin{align}
  \delta_{\phi, \lambda} \mathcal{W} = 0
\end{align}
that reads in detail for $\forall \delta \phi$:
\begin{align}
   0
   &=
   \delta \mathcal{W}
   = 
   2\langle \delta \phi | H-E_0 |  \phi \rangle
   - 
   2\langle \delta \phi | \partial \psi_0 \rangle 
   -
   2\lambda \langle \delta \phi | P_0 | \phi \rangle \; ,
   \label{eq:varW0}
\end{align}
while variation for $\lambda$ recovers the orthogonality condition,
$0=\pd{W}{\lambda}=2\langle \phi | P_0 | \phi \rangle$. 
Eq.~(\ref{eq:varW0}) must hold for any $\delta\phi$, and thus, it is equivalently written as
\begin{align}
  (H-E_0) |\phi\rangle 
  = 
  |\partial\psi_0 \rangle + \lambda P_0 |\phi \rangle \;.
  \label{eq:lineq1lam}
\end{align}
To obtain an expression for $\lambda$, we multiply Eq.~(\ref{eq:lineq1lam}) from the left by
the ground-state electronic wave function, $\langle \psi_0|$, 
\begin{align}
  \langle \psi_0| (H-E_0) |\phi\rangle 
  = 
  \langle \psi_0| \partial\psi_0 \rangle + \lambda \langle \psi_0| P_0 |\phi \rangle  \; 
\end{align}
and obtain after rearrangement
\begin{align}
  \lambda 
  =
  -
  \frac{%
    \langle
      \psi_0|\partial \psi_0
    \rangle
  }{%
    \langle \psi_0 | \phi \rangle
  } \;.
\end{align}
By inserting this expression in Eq.~(\ref{eq:lineq1lam}), we arrive at the linear equation 
\begin{align}
  & (H-E_0) | \phi\rangle 
  = 
  |\partial \psi_0\rangle 
  -
  | \psi_0 \rangle  \langle \psi_0|\partial \psi_0 \rangle \; 
  \nonumber \\
  &\quad\quad\quad
  \ \Leftrightarrow \ 
  (H-E_0) | \phi\rangle 
  = 
  P_0^\perp |\partial \psi_0\rangle  \; ,
  \quad\quad\quad P_0^\perp = 1-P_0
  \label{eq:lineq1}
\end{align}
which can be solved to obtain the $\phi$ perturbed wave function.
Then, $\phi$ is used to compute the mass correction elements according to Eq.~(\ref{eq:rewrite}).
Parameterization of the basis functions used to represent $\phi$ 
can be optimized by minimization of the following functional (instead of the energy functional):
\begin{align}
  & \mathcal{W}
  =
  \langle 
    \phi | H-E_0 | \phi
  \rangle 
  -
  2 \langle \phi | \partial \psi_0 \rangle 
  +
  2\langle \phi | \psi_0 \rangle \langle \psi_0 | \partial \psi_0 \rangle 
  \nonumber \\
  &\quad\quad\quad
  \  \Leftrightarrow\ 
  \mathcal{W}
  =
  \langle 
    \phi | H-E_0 | \phi
  \rangle 
  -
  2 \langle \phi | P_0^\perp | \partial \psi_0 \rangle  \; .
  \label{eq:wval}
\end{align}
It is useful to note, by comparing Eqs.~(\ref{eq:lineq1}) and (\ref{eq:wval}), that
\begin{align}
  \langle \phi | H-E_0 | \phi \rangle 
  =
  \langle \phi | P_0^\perp | \partial \psi_0 \rangle \; .
\end{align}
By inserting this result in Eq.~(\ref{eq:wval}), 
we arrive at 
\begin{align}
  \mathcal{W}
  =
  -\langle \phi | P_0^\perp \partial \psi_0 \rangle \; ,
\end{align}
and this simple expression was used to update the value of $\mathcal{W}$ 
during the course of the basis refinement procedure.

\subsection{Mass correction for electronically excited states \label{ch:ex}}
$\mathcal{W}$ is bounded from below for the ground electronic state, 
but this is not generally true for electronically excited states, 
due to the presence of lower-energy states in the Hamiltonian. 
For the example of the first excited state, we write out $\mathcal{W}$ using the spectral theorem,
\begin{align}
  \mathcal{W}
  =&
  \langle \phi | (H-E_1) | \phi\rangle  - 2 \langle \phi | P_1^\perp \partial \psi_1\rangle  \nonumber \\
  &=
  \sum_{n=0,n\neq 1}^\infty
    (E_n-E_1) \langle \phi | \psi_n \rangle \langle \psi_n | \phi \rangle
  - 2 \langle \phi | P_1^\perp \partial \psi_1\rangle  
  \nonumber \\
  &=
  \underbrace{%
    (E_0-E_1)
    |\langle \phi | \psi_0 \rangle|^2
  }_{\text{negative}}
  +
  \sum_{n>1}^\infty
  \underbrace{%
    (E_n-E_1)
    |\langle \phi | \psi_n \rangle|^2
  }_{\text{non-negative}} 
  - 2 \langle \phi | P_1^\perp \partial \psi_1\rangle
\end{align}
and similarly, for the $k$th electronic state, 
\begin{align}
  \mathcal{W}
  &=
  \langle \phi | (H-E_k)^{-1} | \phi\rangle - 2 \langle \phi | P_k^\perp \partial \psi_k\rangle  \nonumber \\
  &=
  \sum_{i=0}^{k-1}
  \underbrace{%
    (E_i-E_k)
    |\langle \phi | \psi_i \rangle|^2
  }_{\text{negative}}
  +
  \sum_{n>k}^\infty
  \underbrace{%
    (E_n-E_k)
    |\langle \phi | \psi_n \rangle|^2
  }_{\text{non-negative}}  
  - 2 \langle \phi | P_k^\perp \partial \psi_k\rangle  
  \; 
\end{align}
is not bounded from below with respect to the variation of $\phi$. 

We can proceed as follows. First, we exclude the contribution of the (finite many) lower-energy states, next, adapt Schwartz' method to compute the contribution from the (infinitely) many higher-energy states, and in the end, we obtain the total mass correction value by adding the contribution of the $k$ lower-energy states by explicit summation.

To implement this idea, we have to write the functional with $k+1$ auxiliary conditions that ensure the orthogonality of the $\phi^<$ perturbed wave function to all lower-energy states as well as to the $k$th eigenfunction (the superscript $<$ is used to remind ourselves that the perturbed wave function now carries information only about the higher-energy states):
\begin{align}
  \mathcal{W} 
  =
  \langle \phi^< | (H-E_k) | \phi^< \rangle 
  -
  2\langle \phi^< | \partial \psi_k\rangle \; 
\end{align}
with the auxiliary orthogonality conditions
\begin{align}
  \langle \phi^< | P_i | \phi^< \rangle = 0\; ,\quad i = 0, \ldots, k \; 
  \label{eq:auxi}
\end{align}
with $P_i=|\psi_i\rangle \langle\psi_i|$ and we imply during this calculation 
that $\langle \psi_i|\psi_j\rangle=\delta_{ij}$ is fulfilled (the calculation can be generalized 
to non-perfectly orthogonal  electronic states that may occur during numerical computations with different, 
finite basis sets).
We label the excluded space by $\pP=\sum_{i=0}^k P_i$, which equals $P$ defined in Sec.~\ref{ch:coupled},
if the mass correction corresponds to a coupled-state description with all electronic states $i=0,1,\ldots,k$. 
If we describe the $k$th electronic state as an isolated state, then $P=P_k$, whereas $\pP=\sum_{i=0}^k P_i$,
and the contribution of the $0,1,2,\ldots,k-1$ states must be computed by explicit summation.
Further details regarding a coupled-state description including a few electronically excited states are explained
in Sec.~\ref{ch:multi}.

We implement the auxiliary conditions, Eqs.~(\ref{eq:auxi}), using the method of Lagrange multipliers, 
\begin{align}
  \mathcal{W}^\kk[\phi,\lambda_0,\ldots,\lambda_k]
  =
  \langle \phi^< | (H-E_k) | \phi^< \rangle 
  -
  2\langle \phi^< | \partial \psi_k\rangle
  -
  \sum_{i=0}^k 
    \lambda_i 
    \langle \phi^< | P_i | \phi^< \rangle \; . 
  \label{eq:Wmulti0}
\end{align}
Minimization of $\mathcal{W}^\kk$ with respect to the variation of $\phi$ and the $\lambda_i$s assumes the fulfillment of 
the auxiliary orthogonality conditions, Eqs.~(\ref{eq:auxi}), and 
\begin{align}
  0 = 
  2\langle \delta \phi^< | (H-E_k) | \phi^< \rangle
  - 
  2\langle \delta \phi^< | \partial \psi_k\rangle
  -
  2\sum_{i=0}^k 
    \lambda_i 
    \langle \delta \phi^< | P_i | \phi^< \rangle \; , \quad \forall \delta\phi^<   
\end{align}
that is equivalent with
\begin{align}
  (H-E_k) | \phi^< \rangle
  =
  | \partial \psi_k\rangle
  +
  \sum_{i=0}^k 
    \lambda_i 
     P_i | \phi^< \rangle \; .
  \label{eq:lineqmulti}
\end{align}
Similarly to Sec.~\ref{ch:gs}, we obtain $\lambda_j$ by multiplying Eq.~(\ref{eq:lineqmulti})
from the left by the $j$th electronic eigenfunction, $\langle \psi_j|$:
\begin{align}
  \langle 
    \psi_j | (H-E_k) | \phi^< 
  \rangle
  &=
  \langle 
    \psi_j  | \partial \psi_k\rangle
  +
  \sum_{i=0}^k 
    \lambda_i 
      \langle 
    \psi_j | \psi_i\rangle \langle \psi_i | \phi^< \rangle  \;.
\end{align}
Due to $\langle \psi_i|\psi_j\rangle =\delta_{ij}$, the expression simplifies to
\begin{align}
  \langle 
    \psi_j | (H-E_k) | \phi^< 
  \rangle
  =
  \langle 
    \psi_j  | \partial \psi_k\rangle
  +
  \lambda_j
    \langle \psi_j | \phi^< \rangle  
\end{align}
and we obtain
\begin{align}
  \lambda_j 
  =
  \frac{%
    \langle 
      \psi_j | (H-E_k) | \phi^< 
    \rangle
    -
    \langle \psi_j  | \partial \psi_k\rangle    
  }{%
    \langle \psi_j | \phi^< \rangle  
  } \;,\quad j=0,1,\ldots, k \; .
  \label{eq:lambdaj}
\end{align}
As a result, 
\begin{align}
  (H-E_k) | \phi^< \rangle
  &=
  | \partial \psi_k\rangle
  +
  \sum_{i=0}^k 
    \frac{%
      \langle 
        \psi_i | (H-E_k) \phi^< 
      \rangle
      -
      \langle \psi_i  | \partial \psi_k\rangle    
    }{%
      \langle \psi_i | \phi^< \rangle  
    }
    |\psi_i\rangle\langle \psi_i | \phi^< \rangle \nonumber \\
  &=
  | \partial \psi_k\rangle
  +
  \sum_{i=0}^k 
    |\psi_i\rangle   
    \left[%
      \langle 
        \psi_i | (H-E_k) \phi^< 
      \rangle
      -
      \langle \psi_i  | \partial \psi_k\rangle    
    \right]
  \nonumber \\
  &=
  | \partial \psi_k\rangle
  +\pP (H-E_k) |\phi^< \rangle 
  -\pP |\partial \psi_k\rangle    
  \label{eq:lineqmulti1}
\end{align}
that is rearranged to the linear equation 
\begin{align}
  \pP^\perp (H-E_k)|\phi^< \rangle 
  =
  \pP^\perp |\partial \psi_k\rangle  \; ,  
  \label{eq:lineqmulti}
\end{align}
which is solved to obtain the $|\phi^<\rangle$ perturbed wave function (including the effect of all states with an energy higher than
$E_k$).
Using Eqs.~(\ref{eq:Wmulti0}) and (\ref{eq:lambdaj}), 
\begin{align}
  \mathcal{W}^\kk
  &=
  \langle \phi^< | (H-E_k) | \phi^< \rangle 
  -
  2\langle \phi^< | \partial \psi_k\rangle
  \nonumber \\
  &\quad\quad-
  2\sum_{i=0}^k 
  \frac{%
    \langle 
      \psi_i | (H-E_k) \phi^< 
    \rangle
    -
    \langle \psi_i  | \partial \psi_k\rangle    
  }{%
    \langle \psi_i | \phi^< \rangle  
  } 
    \langle \phi^< | P_i | \phi^< \rangle 
  \nonumber \\
  &=
  \langle \phi^< | (H-E_k) | \phi^< \rangle 
  -
  2\langle \phi^< | \partial \psi_k\rangle
  \nonumber \\
  &\quad\quad-
  2\sum_{i=0}^k 
    \langle \phi^< | \psi_i \rangle   
    \left[%
    \langle 
      \psi_i | (H-E_k) \phi^< 
    \rangle
    -
    \langle \psi_i  | \partial \psi_k\rangle    
    \right]
\end{align}
or in short
\begin{align}
  \mathcal{W}^\kk_{\bos{A},\bos{s}}
  &=
  \langle \phi^< | (H-E_k) \pP^\perp | \phi^< \rangle 
  -
  2\langle \phi^< | \pP^\perp \partial \psi_k\rangle
\end{align}
is the (non-linear) functional that can be minimized to optimize the basis function parameterization (that we label with $\bos{A}$ and $\bos{s}$) to systematically improve the perturbed wave function, $|\phi^< \rangle$.
Similarly to Sec.~\ref{ch:gs}, fast evaluation of $\mathcal{W}^\kk_{\bos{A},\bos{s}}$ was carried out by computing
\begin{align}
  \mathcal{W}^\kk_{\bos{A},\bos{s}}
  =
  -\langle \phi^< | \pP^\perp \partial \psi_k\rangle
\end{align}
during the course of the non-linear optimization.

The mass correction for the $\alpha$th isolated, electronically excited state, Eq.~(\ref{eq:massone}), is obtained as
\begin{align}
  M_{\alpha\alpha,ij}/2
  &= 
  \langle \partial_j\psi_\alpha| (\hel-E_\alpha)^{-1} P_\alpha^\perp | \partial_i\psi_\alpha \rangle 
  \nonumber \\
  &=
  \langle \partial_j\psi_\alpha | \pP^\perp | \phi^{(\alpha,i)}_\alpha \rangle 
  +
  \sum_{k=0}^\alpha 
    \frac{%
      \langle 
        \partial_j\psi_\alpha | \psi_k
      \rangle 
      \langle 
        \psi_k | \partial_i\psi_\alpha
      \rangle 
    }{E_k-E_\alpha}
  \nonumber \\
  &=
  \langle 
    \phi^{(\alpha,j)}_\alpha 
    | \pP^\perp | 
    \partial_i\psi_\alpha 
  \rangle 
  +
  \sum_{k=0}^\alpha 
    \frac{%
      \langle 
        \partial_j\psi_\alpha | \psi_k
      \rangle 
      \langle 
        \psi_k | \partial_i\psi_\alpha
      \rangle 
    }{E_k-E_\alpha}  \; ,
  \label{eq:massex}
\end{align}
where $\phi^{(\alpha,i)}_\alpha$ and $\phi^{(\alpha,j)}_\alpha$ 
correspond to the $\phi$ perturbed wave function computed with 
with $\partial=\partial_i$ and $\partial=\partial_j$, respectively.
In relation with Eq.~(\ref{eq:massex}), it is important to emphasize that 
$P_\alpha=|\psi_\alpha\rangle \langle\psi_\alpha|$, whereas
$\pP=\sum_{k=0}^\alpha P_k$.

\subsection{Mass correction for coupled electronic manifolds \label{ch:multi}}
The ideas outlined for the excited state computation can be straightforwardly used and implemented 
for a coupled electronic subspace, 
\begin{align}
  P_\cpl = \sum_{n={\minC}}^{\maxC} |\psi_n\rangle \langle \psi_n| \; .
\end{align}
$C$ labels the set of the indices of the electronic states that are included in $P_C$.
We will assume that the electronic states are numbered in an increasing energy order, \emph{i.e.,} $\psi_{\maxC}$ is the highest-energy state in the coupled subspace.
The general form of the mass-correction coupling between the $\alpha$th and $\beta$th states from $P_\cpl$, Eq.~(\ref{eq:massmulti}) (we work in
an adiabatic basis), is
\begin{align}
  M_{\alpha\beta, ij}
  &=
  \langle
    \partial_{j} \psi_\alpha | \left[(H-E_\alpha)^{-1}+(H-E_\beta)^{-1}\right] P_{\cpl}^\perp | \partial_{i} \psi_\beta
  \rangle 
  \nonumber \\
  &=
  \sum_{n\not\in \cpl}
  \frac{%
    \langle
      \partial_{j} \psi_\alpha | \psi_n \rangle \langle \psi_n | \partial_{i} \psi_\beta 
    \rangle
  }{E_n-E_\alpha}
  +
  \sum_{n\not\in \cpl}
  \frac{%
  \langle
    \partial_{j} \psi_\alpha | \psi_n \rangle \langle \psi_n | \partial_{i} \psi_\beta
  \rangle
  }{E_n-E_\beta} \; .
  \label{eq:cplmmx}
\end{align}
If $\alpha\neq\beta$, two perturbed wave functions are computed. For example,
\begin{align}
  \langle \phi_{\alpha}^{(j,\alpha)}|
  =
  \sum_{n > \maxC}
  \frac{%
    \langle 
      \partial_j \psi_\alpha |
      \psi_n \rangle \langle \psi_n | 
  }{E_n-E_\alpha}
\end{align}
and
\begin{align}
  |\phi_{\beta}^{(i,\beta)}\rangle
  =
  \sum_{n > \maxC}
  \frac{%
    | \psi_n \rangle \langle \psi_n | \partial_{i} \psi_\beta
  \rangle
  }{E_n-E_\beta} \; ,   
\end{align}
where the infinite sums are not computed explicitly, but the perturbed wave functions (left hand side of the equations)
are obtained by using the generalized Schwartz' method (Secs.~\ref{ch:gs} and \ref{ch:ex}).
Then, the coupled-state mass matrix element is obtained as
\begin{align}
  M_{\alpha\beta;ij}
  &=
  \sum_{n=0}^{\maxC}
    \frac{%
      \langle\partial_j\psi_\alpha|\psi_n\rangle \langle\psi_n |\partial_i\psi_\beta \rangle
    }{%
      E_n-E_\alpha
    }  
  +
  \langle
    \phi^{(j,\alpha)}_\alpha 
    |\pP_\cpl^\perp |
    \partial_i \psi_\beta
  \rangle  
  \nonumber \\
  & 
  +
  \sum_{n=0}^{\maxC}
    \frac{%
      \langle\partial_j\psi_\alpha|\psi_n\rangle \langle\psi_n |\partial_i\psi_\beta \rangle
    }{%
      E_n-E_\beta
    }  
  +
  \langle
    \partial_j \psi_\alpha |\pP_\cpl^\perp | 
    \phi^{(i,\beta)}_\beta
  \rangle   
  \; .
\end{align}

\section{Implementation and computational details}
The theoretical approach outlined in Sec.~\ref{ch:method} has been implemented in the in-house developed computer program named QUANTEN (QUANTum mechanical description of Electrons and atomic Nuclei). QUANTEN has recent applications including non-relativistic energy upper and lower bounds, non-adiabatic, pre-Born--Oppenheimer, perturbative and variational relativistic computations 
\cite{FeMa22H3,IrJeMaMaRoPo21,Ma18nonad,Ma18he2p,FeMa19HH,Ma19rev,FeMa19EF,MaCa21,FeKoMa20,JeIrFeMa21,JeFeMa22,FeJeMa22}. 
The program contains a (stochastic and deterministic) non-linear variational engine and an integral library for variants of explicitly correlated Gaussian (ECG) functions. The electronic wave function is written as a linear combination of anti-symmetrized products of spatial and spin functions.
In this work, the spatial functions are floating ECGs, 
\begin{align}
   \varphi(\bos{r},\bos{A},\bos{s})
   =
   \exp\left[%
     -(\bos{r}-\bos{s})\Tt (\bos{A}\otimes \bos{I}_3) (\bos{r}-\bos{s})
   \right] \; ,
\end{align}
where 
$\bos{r}\in\mathbb{R}^{6}$ collects the Cartesian coordinates of the two electrons, 
$\bos{A}\in\mathbb{R}^{2\times 2}$ is a symmetric, positive-definite parameter matrix, 
and the $\bos{s}\in\mathbb{R}^6$ shift vectors are fixed to the proton-proton axis. 
The functions are adapted to have gerade (g) symmetry,
and as a result of this construct, the spatial basis functions have $\Sigma_\text{g}^+$ symmetry.

The \XX, \EF, \GK, \HH, \Sfive, and \Ssix\ electronic states are computed with this setup using 1200 ECG functions optimized separately for each electronic state state. 
First, the electronic energy is converged
at a single point (1.4~bohr for the \XX\ state and 3~bohr for the \EF, \GK, and \HH\ states) within a few n\Eh\ precision. 
Then, 
a series of points is generated by rescaling the centers ($\bos{s}$ vectors) upon changing the distance (by $\Delta R=0.1$~bohr at every step) using the rescaling procedure proposed by Cencek and Kutzelnigg \cite{CeKu1997}. After rescaling, we have performed repeated refinement cycles at every new geometry. At a few selected points, the \Sfive\ and \Ssix\ states were computed by running  repeated energy refinement cycles for the minimization of the \Sfive\ and \Ssix\ energy starting from the basis set optimized for the \HH\ state at the same geometry. The resulting \Sfive\ and \Ssix\ energies are converged within 50-100~n\Eh, and this value can be reduced to 10~n\Eh\ in the present setup without major computational effort.

The wave function derivatives, $| \partial \psi_\alpha\rangle$ with respect to the nuclear coordinates have been computed by finite differences and the rescaling procedure of Ref.~\cite{CeKu1997}. Instead of the six Cartesian coordinates of the two protons, we fix the protons center of mass at the origin and use spherical polar coordinates ($R,\vtheta,\phi$) to describe the shape and orientation of the molecule. Hence, for the vibronic mass computations, we had to compute only the $\partial/\partial R$, henceforth $\partial$,
derivative of the electronic states. The $\partial \psi_\alpha$ function, as well as the $\phi$ perturbed wave function (Sec.~\ref{ch:method}) have $\Sigma_\text{g}^+$ symmetry. For this reason, the basis set optimized for the electronic state was usually an excellent starting basis for the $\phi$ perturbed wave function, and a few refinement steps were performed. 

In this initial report, the vibronic mass-correction values have been computed for 26 nuclear configurations (see \som). The non-adiabatic coupling and the diagonal and off-diagonal Born--Oppenheimer corrections, Eqs.~(\ref{eq:cplham})--(\ref{eq:cplkeo}), are taken from the work of Wolniewicz and Dressler \cite{WoDr94}, although the Born--Oppenheimer (BO) potential energies are replaced with the PECs computed in the present work. We also note that BO potential energy curves have been recently reported in the literature with a $10^{-10}$ relative precision \cite{SiZsPa18} that will be very useful when all other contributions (non-adiabatic, relativistic, and QED) will have been checked and refined. 

It was necessary to check the phase of the wave functions ($\partial \psi_\alpha$) in the bra and in the ket for the off-diagonal elements of the mass-correction tensor. All phases were adjusted with respect to the phases at $R=3$~bohr, and they were adjusted to be identical with the phase corresponding to the non-adiabatic coupling matrix elements taken from Ref.~\cite{WoDr94} that was checked at single points.

The physical constants and conversion factors used in the computations
were taken from the CODATA18 recommendation,
$m_\alpha / m_\text{e} = 1\,836.152\,673\,43(11)$
and 
$1~\text{E}_\text{h} =219\,474.631\,363\,20(43)$~\cm. \\



\begin{figure}
%
\includegraphics[scale=1.1]{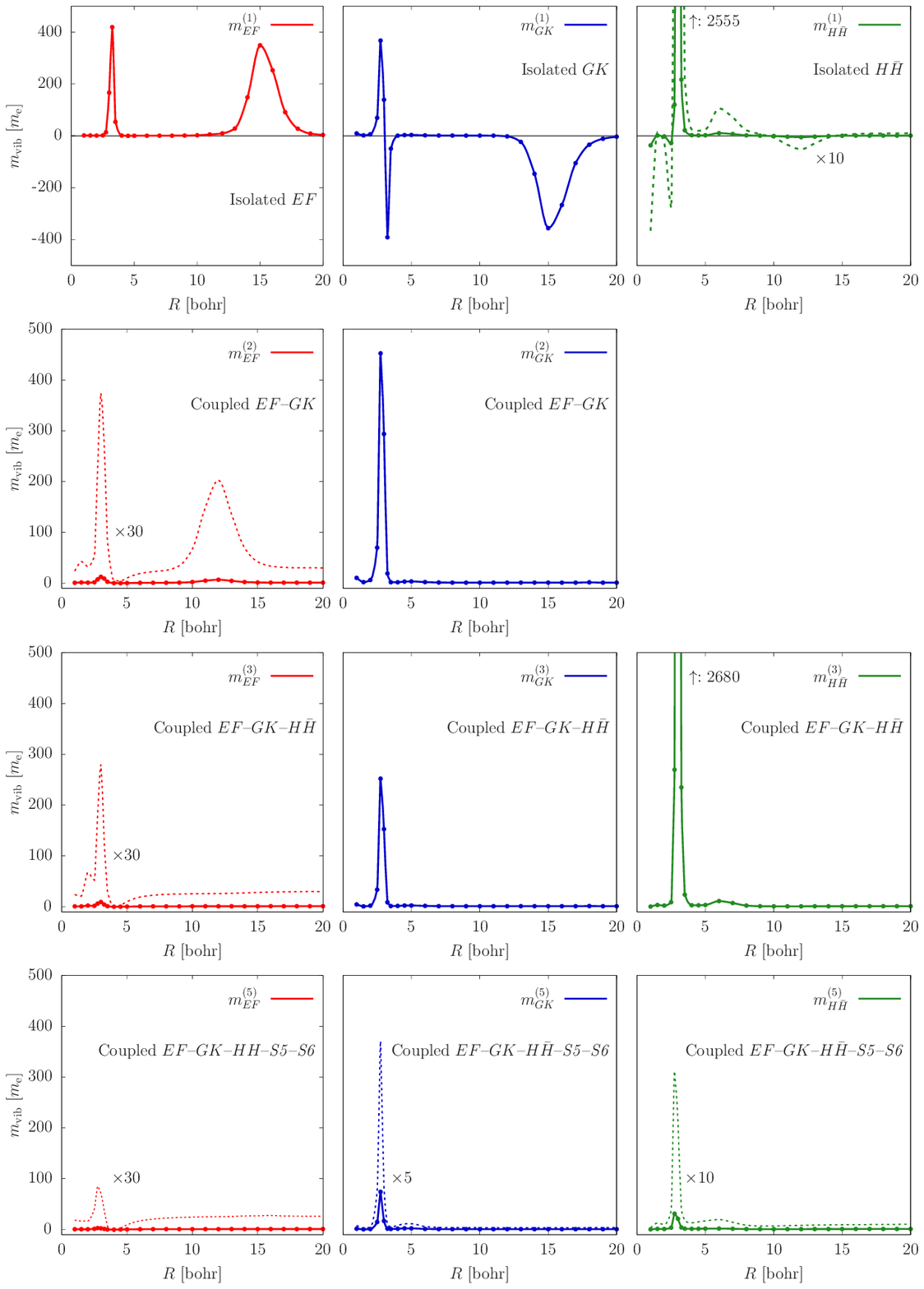} 
\caption{%
  Overview of the diagonal vibronic mass correction values for the \EF, \GK, and \HH\ electronic states
  corresponding to a one- (1: \EF/\GK/\HH), two- (2: \EF--\GK), three- (3: \EF--\GK--\HH),
  and five-dimensional (5: \EF--\GK--\HH--\Sfive--\Ssix) coupled electronic manifold. 
  Small features can be better observed in the $n$-fold enlargement ($\times n$) of the functions plotted in dashed line.
  \label{fig:vibmass}
}
\end{figure}

\begin{figure}
\includegraphics[scale=1]{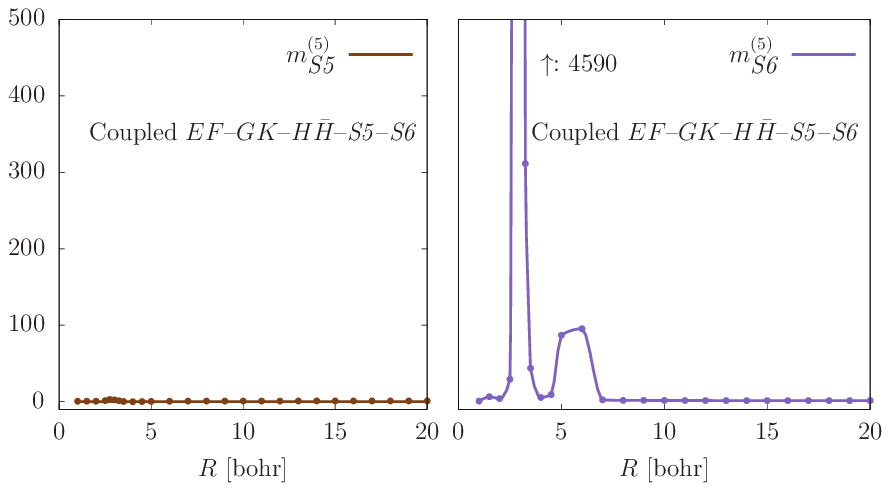}  
\caption{%
  Diagonal vibronic mass correction values for the \Sfive\ and \Ssix\ electronic states
  described within the five-dimensional (5: \EF--\GK--\HH--\Sfive--\Ssix) coupled electronic manifold. 
  \label{fig:mvib2}
}
\end{figure}

\section{Vibronic computations for the hydrogen molecule \label{ch:vibronic}}
\subsection{Vibronic masses \label{ch:vibmass}}
Figures~\ref{fig:vibmass} and \ref{fig:mvib2} show the vibronic mass correction values (corresponding to the $R$ degree of freedom \cite{Ma18nonad}) computed in the present work for the \EF, \GK, and \HH\ electronic states corresponding to a single state description (1-dimensional electronic subspace), and for the \EF--\GK\ (2), the \EF--\GK--\HH\ (3), and the \EF--\GK--\HH--\Sfive--\Ssix\ (5) multi-dimensional descriptions. 
The off-diagonal vibronic mass correction values 
and all numerical data (points) used to prepare the figures are deposited in the \som.
Comparison of the computed vibronic energies and experiment is shown in Fig.~\ref{fig:exptcomp}.
During the discussion of the results, it may be relevant to inspect also Fig.~\ref{fig:pec} that provides an overview of the relevant electronic states. 

The vibronic mass correction is always positive for the ground state (diagonal elements), but it can also be negative for excited states. 
We see large negative features in 
the isolated \GK\ state due to the nearby, lower-energy \EF\ state, and it is interesting to note the corresponding (positive) feature in the isolated \EF\ vibronic mass correction curve. 
Nevertheless, these features appear to be of purely theoretical interest, since 
the single-state description of the \EF\ or the \GK\ state does not give good results (of spectroscopic quality). For these single-state computations, the second-order perturbative correction is insufficient for an accurate description. Regarding higher-order corrections, already the third-order corrections \cite{MaTe19} appear to be numerically very complicated. 

In contrast to the \EF\ and the \GK\ states, for \emph{the outer well} of the \HH\ state (Fig.~\ref{fig:pec}), the single-state adiabatic description was known to give reasonable results and was found to be useful in terms of the assignment of the spectrum \cite{ReHoUbWo99}.
Over this potential energy well, we observe a shallow (negative) minimum of the vibronic mass correction ($m_{H\bar{H}}^{(1)}$ in Fig.~\ref{fig:vibmass}), \emph{i.e.,} the effective vibronic mass is smaller than the mass of the proton. This feature was computed already in Ref.~\cite{FeMa19HH}, and it was found that correction of the constant, nuclear (proton) mass by this non-adiabatic term the deviation of theory and experiment is reduced by an order of magnitude, \emph{i.e.,} from ca. 1~\cm\ to ca. 0.1~\cm . (We note that in both the single-state adiabatic and non-adiabatic computations relativistic and leading-order QED corrections were included in Ref.~\cite{FeMa19HH}.)

For further improvement, it would be necessary  either to account for higher-order perturbative corrections, which is numerically very complicated, or to replace the 1-dimensional electronic subspace with a multi-dimensional subspace by including the nearby-lying electronic states in the coupled electronic subspace that is feasible and subject of the present work.

\section{Vibronic energies}
Following Wolniewicz, Dressler and their co-workers \cite{WoDr77,DrGaQuWo79,QuDrWo90,WoDr94,YuDr94}
the non-adiabatic wave function is expanded as
\begin{align}
  \Psi
  =
  \sum_{\alpha\in C} 
    \psi_\alpha(r,R) \frac{1}{R} f_\alpha(R) 
  =
  \frac{1}{R} \bos{\psi} \bos{f} \; ,
\end{align}
and the vibrational part of the wave function satisfies
\begin{align}
  \left\lbrace%
    -\frac{1}{2\mu}    
    \left[%
      \frac{\dd}{\dd R}
        \left(%
          \bos{I}   
          - 
          \frac{1}{2\mu}\bos{m}
        \right)
      \frac{\dd}{\dd R}
      + \bos{A} 
      + \bos{B} \frac{\dd}{\dd R}
    \right]
    + \bos{U}
  \right\rbrace
  \bos{f}
  =
  E\bos{f} \; ,
  \label{eq:coupledeq}
\end{align}
where $\mu=m_\text{p}/2$ is the reduced mass, $\bos{I}_{\alpha\beta}=\delta_{\alpha\beta}$, and the $A_{\alpha\beta}$ and $B_{\alpha}$ coupling functions were taken from Ref.~\cite{WoDr94}. 
We use the adiabatic representation, hence $(\bos{U})_{\alpha\beta}=\delta_{\alpha\beta} V_\alpha$ is diagonal.
The vibronic mass correction elements, $m_{\alpha\beta}$, and the BO PECs, $V_\alpha$, were computed in the present work. 
Simple truncation of the electronic space, Eqs.~(\ref{eq:cplham})--(\ref{eq:cplkeo}), corresponds to neglecting $\bos{m}$ in Eq.~(\ref{eq:coupledeq}), whereas solving the complete Eq.~(\ref{eq:coupledeq}) corresponds to the second-order effective Hamiltonian, Eq.~(\ref{eq:nadham}).

We note that $\bos{m}$ corresponds to the matrix representation over the $\psi_\alpha \in P_C$ electronic eigenfunctions (adiabatic representation) of the $R,R$ element of the mass-correction tensor, Eq.~(\ref{eq:massmultiad}), expressed in spherical polar coordinates \cite{Ma18nonad}.

To solve Eq.~(\ref{eq:coupledeq}), we used the associated Laguerre polynomials, $L_n^{(\alpha)}$ with $\alpha=2$ and the discrete variable representation (DVR) \cite{LiCa00} for every $f_\alpha(R)$ function similarly to Refs.~\cite{Ma18nonad,Ma18he2p}.

\vspace{0.5cm}
Although the present computations do not contain relativistic and QED corrections, comparison of the vibronic term values with experiment is relevant, because we think that non-adiabatic effects have an important role
in the earlier deviation of theory and experiment \cite{YuDr94}, which is larger than 1~\cm\ for several states.
For this comparison, we have calculated the non-relativistic term value, $T=E-E_{{X0},\text{nr}}$, where $E_{{X0},\text{nr}}$ is the non-relativistic, non-adiabatic energy of the rovibronic ground state ($X0$). We use $E_{{X0},\text{nr}}=-1.164\ 025\ 031$~\Eh\ \cite{PaKo09} that is sufficiently precise for this work, but we note that further digits are available \cite{PaKo18}.
Regarding the experimental values, we use the dataset from Ref.~\cite{BaSaVeUb10}, but we also note that for some of the terms corresponding to the \EF\ \cite{DiSaNiJuRoUb12}, the \GK\ and the \HH\ (inner well) states \cite{HoBeMe18} more precise experimental data has became available since Ref.~\cite{BaSaVeUb10} that is beyond the current theoretical accuracy.

Direct comparison of the computed non-adiabatic energies would be most appropriate with pre-Born--Oppenheimer (preBO, here: four-particle) energies, which do not contain relativistic and QED effects. In the $^1\Sigma_\text{g}^+$ manifold, the preBO energy is  available only for the vibrational ground state \emph{(}\emph{E0} and its rotational excitations) corresponding to the \EF\ electronic state \cite{FeMa19EF}. For the \emph{E0} state the comparison is shown in Table~\ref{tab:prebonad}.

\begin{table}
\caption{%
  Comparison of the four-particle, pre-Born--Oppenheimer (preBO) \cite{FeMa19EF} and non-adiabatic (nad) energies (this work)  for the lowest vibrational level corresponding to the \EF\ \Sgp\ electronic state.
  \label{tab:prebonad}
}
\begin{tabular}{@{}l@{\ \ \ }l@{\ \ \ }r@{}}
\hline\hline\\[-0.35cm]
  Coupled states & 
  Mass$^\ast$ & 
  $T_\text{preBO}-T_\text{nad}$ [cm$^{-1}$] \\
  \hline\\[-0.35cm]
  \EF--\GK--\HH                & $m_\text{p}$   & $-$0.28 \\
  \EF--\GK--\HH                & $m_\text{eff}$ &    0.14 \\
  \EF--\GK--\HH--\Sfive--\Ssix & $m_\text{p}$   & $-$0.27 \\
  \EF--\GK--\HH--\Sfive--\Ssix & $m_\text{eff}$ & $-$0.05\\
\hline\hline\\[-0.35cm]
\end{tabular}
  ~\\
  $^\ast$: $m_\text{p}$ and $m_\text{eff}$ refer to Eq.~(\ref{eq:coupledeq})
  without and with the $\bos{m}$ vibronic mass correction term, respectively.
\end{table}

\begin{figure}
\scalebox{1.}{%
\begin{tabular}{@{}cc@{}}
  \includegraphics[scale=1.]{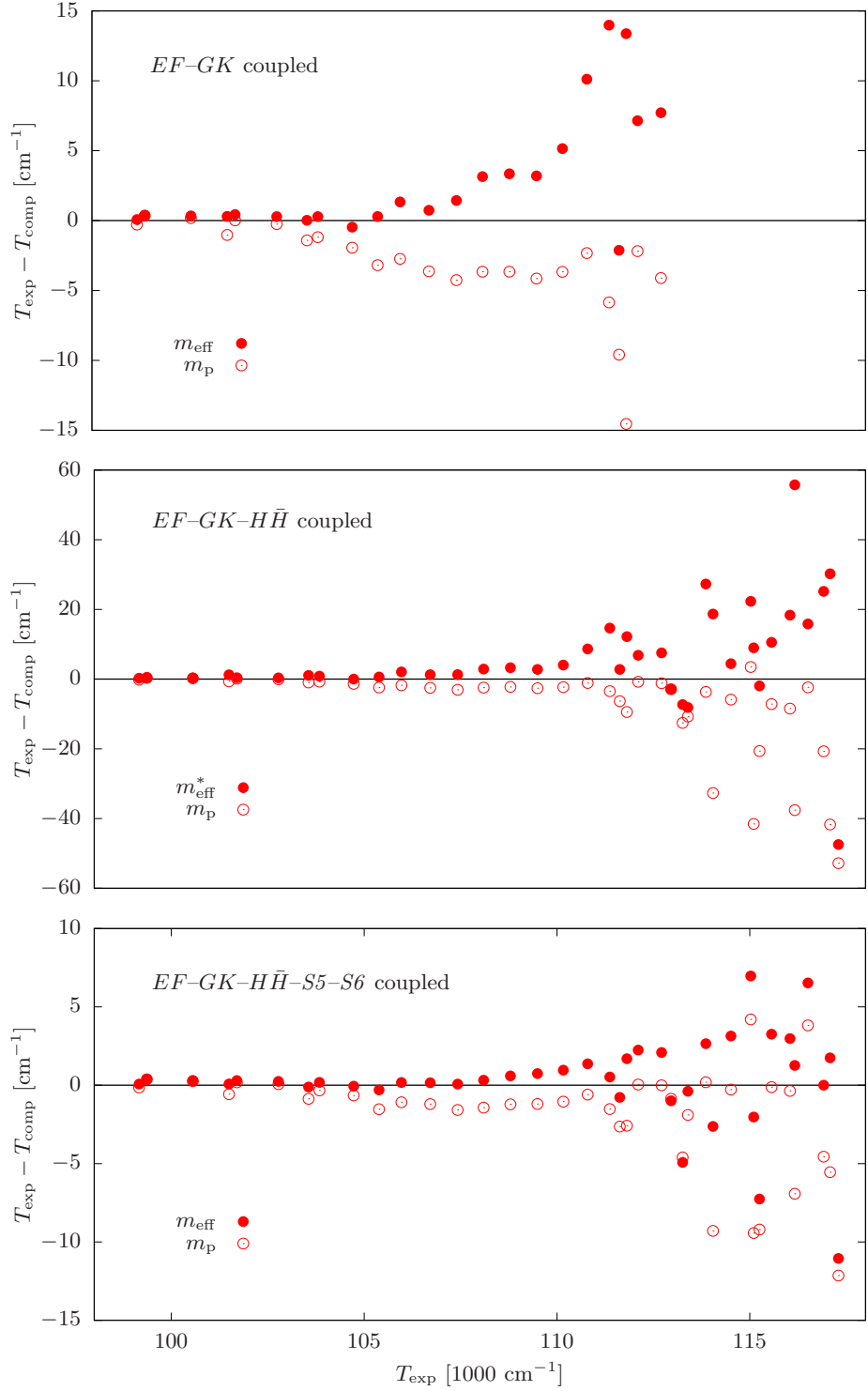}
\end{tabular}
}
\caption{%
  Deviation of the experimental \cite{BaSaVeUb10} and computed term values corresponding to the electronically excited \Sgp\ manifold of molecular hydrogen. 
  $T_\text{comp}=E-E_{X0,\text{nr}}$ with the $E_{X0,\text{nr}}=-1.164\ 025\ 031$~\Eh\ \cite{PaKo09} non-relativistic energy of the vibronic ground state ($X0$). 
  The computed $E$ energies are obtained from explicit coupling of 2 (\EF--\GK), 3 (\EF--\GK--\HH), and 5 (\EF--\GK--\HH--\Sfive--\Ssix) electronic states 
  using the nuclear mass ($m_\text{p}$)
  or 
  including also the vibronic mass correction ($m_\text{eff}$)
  that corresponds to solving 
  Eq.~(\ref{eq:coupledeq}) without or with the $\bos{m}$ correction term, respectively.
  In the \EF--\GK\ subfigure no datapoints are shown beyond the fundamental \HH\ vibrational energy.
  \label{fig:exptcomp}
}
\end{figure}

Figure~\ref{fig:exptcomp} shows the deviation of the experimental \cite{BaSaVeUb10} and non-relativistic, non-adiabatic term values corresponding to coupling 2 (\EF--\GK), 3 (\EF--\GK--\HH), and 5 (\EF--\GK--\HH--\Sfive--\Ssix) electronic states. Before inclusion of the relativistic and QED corrections, which would allow direct comparison of theory and experiment, further improvements to the current computations will be necessary. The computation of the non-relativistic Bethe logarithm, which appears in the leading-order QED corrections, is in progress \cite{FeMa22bethe}. 
At the present stage, we may observe that without the non-adiabatic mass corrections, the computed term values overestimate the experimental values, whereas when the mass corrections are included the computed non-relativistic term value is typically larger in this range than the experimental term energy. The preBO study of the \EF\ levels \cite{FeMa19EF} shows that the relativistic and QED corrections to the term value is negative, which suggests that the current inclusion of the non-adiabatic masses does improve upon the truncated (proton mass) results (see also Table~\ref{tab:prebonad}).

For higher-energy states, we observe that inclusion of the effective vibronic masses can shift the term values by several (tens of) wave numbers to lower energies, which correspond to a positive shift in the energy levels by values as large as  5--20~\cm.
It is necessary to note that a large shift can indicate that
(a) second-order perturbation theory is insufficient (and we would need to include higher-order terms); 
or (b) the actively coupled electronic space is too small, or in other words, the energy gap between the coupled space and the discarded states is too small.

\begin{figure}
  \includegraphics[scale=0.6]{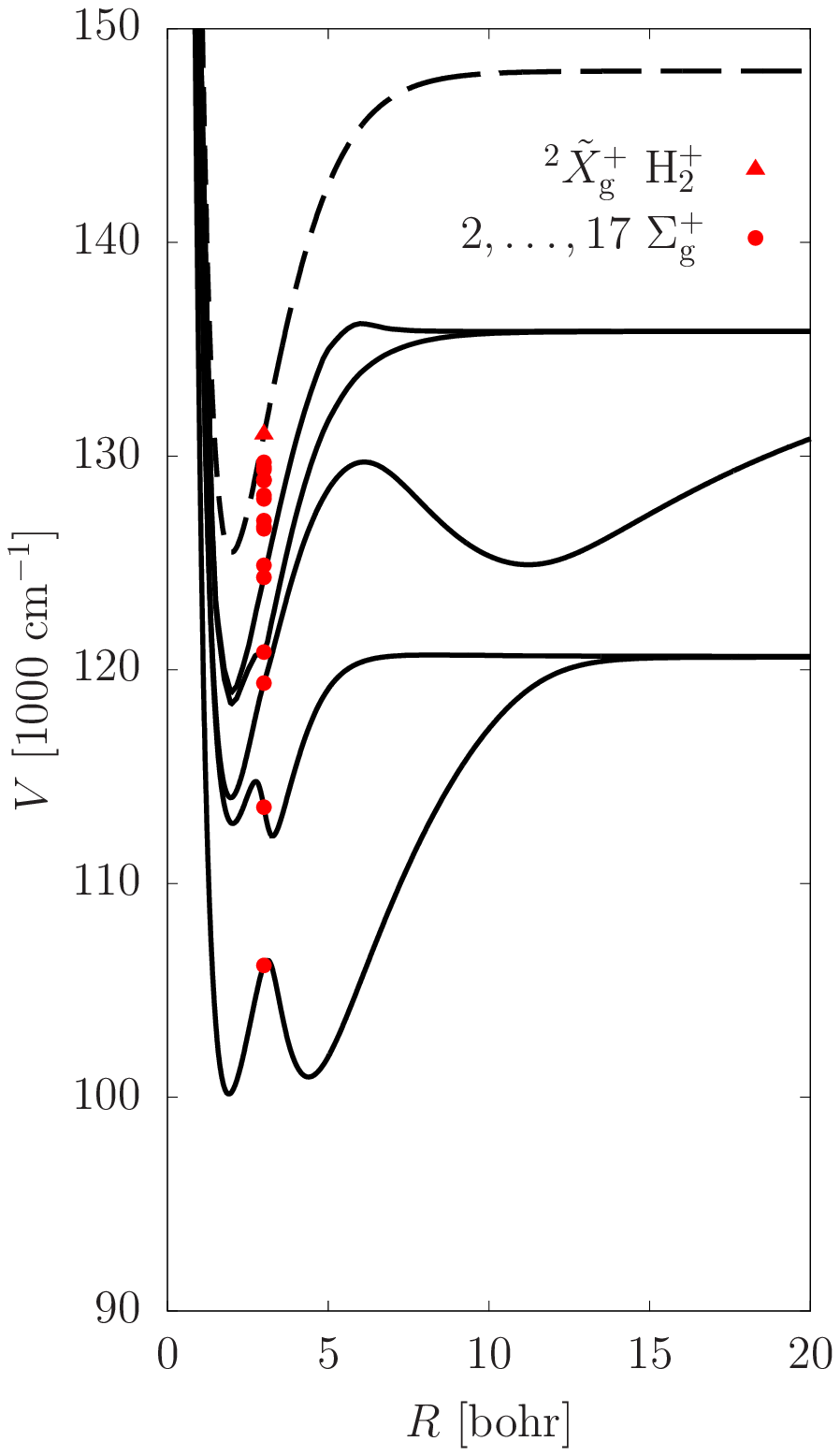}
  \caption{%
    The \EF, \GK, \HH, \Sfive, and \Ssix\ \Sgp\ BO potential energy curves considered in this work. 
    The red points highlight the $n$~\Sgp\ with $n=1,\ldots,17$ potential energy values 
    that we computed below the $^2\tilde{X}_\text{g}^+$ ground-state energy of H$_2^+$ at $R=3$~bohr internuclear separation.
    The potential energy curve of H$_2^+$ was taken from Ref.~\cite{BeMe16b}. 
    \label{fig:pec}
  }
\end{figure}

The third-order terms \cite{MaTe19} appear to be too complicated (at the moment) for numerical evaluation, while, enlargement of the coupled space is feasible, although it raises some further (fundamental) questions for the theory (see Sec.~\ref{ch:end} and Fig.~\ref{fig:pec}).

How can we improve upon the current results results? 

There is ample space for improvements. First of all, it will be necessary to recompute the coupling functions, $\bos{A}$ and $\bos{B}$ in Eq.~(\ref{eq:coupledeq}), more precisely, and to compute the non-adiabatic mass corrections at more points along the PECs. 
It also appears to be necessary to add a few more electronic states to the actively coupled manifold that should also be computationally feasible. 

It is necessary to note however that if a higher-energy actively coupled electronic state is close to another electronic state not included in $P_C$, then it can have a large mass-correction value (that may cause computational instabilities), while its contribution to the dynamics in the interesting energy range may be small. 
We have experienced this problem for the $m_{\HH}^{(3)}$ value in the \EF--\GK--\HH\ computation, and we handled it with an \emph{ad hoc} numerical damping. So, in practice,  instead of 2680, we used 500 at the maximum of the $m_{\HH}^{(3)}$ peak near $R=3$~bohr (Fig.~\ref{fig:vibmass}). This makes the numerical computations more stable, but it introduces some uncertainty (ca. 0.02--0.1~\cm) in the higher energy range.

So, the question arises: how many more electronically excited states do we need to couple to have precise energies for the \EF--\GK--\HH\ manifold? Is the gap condition of perturbation theory~\cite{Te03} well fulfilled in practice, \emph{i.e.,} is it possible to choose the subspace so that there is a sufficiently large gap of the coupled states and the rest? In other words, is the second-order effective Hamiltonian, Eq.~(\ref{eq:nadham}), sufficient to obtain accurate rovibronic energies? 
Regarding higher excited PECs of H$_2$ (from the $^1\Sigma_\text{g}^+$ manifold), 
Corongiu and Clementi computed many states, including many $^1\Sigma_\text{g}^+$ states, of the H$_2$ molecule with a ca. $10^{-5}$~\Eh\ precision \cite{CoCl09,CoCl10}.
Excited $n\Sigma_\text{g}^+$ states up to $n=7$ have been computed with a relative precision of $10^{-10}$ (0.000\ 02 \cm) by
Si\l kowski, Zientkiewicz, and Pachucki \cite{SiZsPa18}.
Figure~\ref{fig:pec} shows the PECs used in this work and the electronic energy of the $n\Sigma_\text{g}^+$ states at $R=3$~bohr proton-proton distance up to the BO energy of the $^2{\tilde X}^+\text{g}$ state of H$_2^+$. At $R=3$~bohr, 
we have found 17 electronic states below the $^2{\tilde X}^+_\text{g}$ energy value, which indicates that in this region, \emph{i.e.,} beyond 124 000~\cm\ up to ionization,  there is a very high density of states, ca. 556~\cm\ per state.

\section{Summary, conclusion, and outlook \label{ch:end}}

A variational approach has been reported for the computation of the non-adiabatic mass correction matrix that includes a sum-over-states expression.
The approach can be used for isolated ground or electronically excited states, as well as for coupled electronic subspaces.

Initial results have been reported about of the vibronic mass correction for the \EF--\GK--\HH--\Sfive--\Ssix\~\Sgp\ manifold and the effect of the mass correction term on the vibronic energies. For a direct comparison with experiment, it is necessary to account for the relativistic and leading-order QED corrections, for which the computation of the Bethe logarithm is in progress \cite{FeMa22bethe}. 

In future work, it will be necessary to 
(a) compute more precise coupling matrix elements, 
(b) compute the mass correction values at more points, and 
(c) possibly couple more electronic states. At
the same time, the high density of electronic states beyond the $H\bar{H}$ state (Fig.~\ref{fig:pec}) may require further development of the theory. 


\vspace{0.5cm}
\begin{acknowledgments}
\noindent 
EM thanks Stefan Teufel for discussions about adiabatic perturbation theory.
Financial support of the European Research Council through a Starting Grant (No.~851421) is gratefully acknowledged. 
DF thanks a doctoral scholarship from the ÚNKP-21-3 New National Excellence Program of the Ministry for Innovation and Technology from the source of the National Research, Development and Innovation Fund (ÚNKP-21-3-II-ELTE-41).
\end{acknowledgments}

\clearpage

\begin{thebibliography}{44}%
\makeatletter
\providecommand \@ifxundefined [1]{%
 \@ifx{#1\undefined}
}%
\providecommand \@ifnum [1]{%
 \ifnum #1\expandafter \@firstoftwo
 \else \expandafter \@secondoftwo
 \fi
}%
\providecommand \@ifx [1]{%
 \ifx #1\expandafter \@firstoftwo
 \else \expandafter \@secondoftwo
 \fi
}%
\providecommand \natexlab [1]{#1}%
\providecommand \enquote  [1]{``#1''}%
\providecommand \bibnamefont  [1]{#1}%
\providecommand \bibfnamefont [1]{#1}%
\providecommand \citenamefont [1]{#1}%
\providecommand \href@noop [0]{\@secondoftwo}%
\providecommand \href [0]{\begingroup \@sanitize@url \@href}%
\providecommand \@href[1]{\@@startlink{#1}\@@href}%
\providecommand \@@href[1]{\endgroup#1\@@endlink}%
\providecommand \@sanitize@url [0]{\catcode `\\12\catcode `\$12\catcode
  `\&12\catcode `\#12\catcode `\^12\catcode `\_12\catcode `\%12\relax}%
\providecommand \@@startlink[1]{}%
\providecommand \@@endlink[0]{}%
\providecommand \url  [0]{\begingroup\@sanitize@url \@url }%
\providecommand \@url [1]{\endgroup\@href {#1}{\urlprefix }}%
\providecommand \urlprefix  [0]{URL }%
\providecommand \Eprint [0]{\href }%
\providecommand \doibase [0]{https://doi.org/}%
\providecommand \selectlanguage [0]{\@gobble}%
\providecommand \bibinfo  [0]{\@secondoftwo}%
\providecommand \bibfield  [0]{\@secondoftwo}%
\providecommand \translation [1]{[#1]}%
\providecommand \BibitemOpen [0]{}%
\providecommand \bibitemStop [0]{}%
\providecommand \bibitemNoStop [0]{.\EOS\space}%
\providecommand \EOS [0]{\spacefactor3000\relax}%
\providecommand \BibitemShut  [1]{\csname bibitem#1\endcsname}%
\let\auto@bib@innerbib\@empty
\bibitem [{\citenamefont {Wolniewicz}\ and\ \citenamefont
  {Dressler}(1977)}]{WoDr77}%
  \BibitemOpen
  \bibfield  {author} {\bibinfo {author} {\bibfnamefont {L.}~\bibnamefont
  {Wolniewicz}}\ and\ \bibinfo {author} {\bibfnamefont {K.}~\bibnamefont
  {Dressler}},\ }\bibfield  {title} {\bibinfo {title} {The {EF} and {GK} states
  of hydrogen: {A}diabatic calculation of vibronic states in {H}$_2$, {HD}, and
  {D}$_2$},\ }\href {https://doi.org/10.1016/0022-2852(77)90050-9} {\bibfield
  {journal} {\bibinfo  {journal} {J. Mol. Spectrosc.}\ }\textbf {\bibinfo
  {volume} {67}},\ \bibinfo {pages} {416} (\bibinfo {year} {1977})}\BibitemShut
  {NoStop}%
\bibitem [{\citenamefont {Dressler}\ \emph {et~al.}(1979)\citenamefont
  {Dressler}, \citenamefont {Gallusser}, \citenamefont {Quadrelli},\ and\
  \citenamefont {Wolniewicz}}]{DrGaQuWo79}%
  \BibitemOpen
  \bibfield  {author} {\bibinfo {author} {\bibfnamefont {K.}~\bibnamefont
  {Dressler}}, \bibinfo {author} {\bibfnamefont {R.}~\bibnamefont {Gallusser}},
  \bibinfo {author} {\bibfnamefont {P.}~\bibnamefont {Quadrelli}},\ and\
  \bibinfo {author} {\bibfnamefont {L.}~\bibnamefont {Wolniewicz}},\ }\bibfield
   {title} {\bibinfo {title} {The {EF} and {GK} {$^1\Sigma_\mathrm{g}^+$}
  states of hydrogen: {C}alculation of nonadiabatic coupling},\ }\href
  {https://doi.org/10.1016/0022-2852(79)90117-6} {\bibfield  {journal}
  {\bibinfo  {journal} {J. Mol. Spectrosc.}\ }\textbf {\bibinfo {volume}
  {75}},\ \bibinfo {pages} {205} (\bibinfo {year} {1979})}\BibitemShut
  {NoStop}%
\bibitem [{\citenamefont {Senn}\ \emph {et~al.}(1988)\citenamefont {Senn},
  \citenamefont {Quadrelli},\ and\ \citenamefont {Dressler}}]{SeQuDr88}%
  \BibitemOpen
  \bibfield  {author} {\bibinfo {author} {\bibfnamefont {P.}~\bibnamefont
  {Senn}}, \bibinfo {author} {\bibfnamefont {P.}~\bibnamefont {Quadrelli}},\
  and\ \bibinfo {author} {\bibfnamefont {K.}~\bibnamefont {Dressler}},\
  }\bibfield  {title} {\bibinfo {title} {The {$B\ ^1\Sigma_\mathrm{u}^+$},
  {$B'\ ^1\Sigma_\mathrm{u}^+$}, {$C\ ^1\Pi_\mathrm{u}$}, and {$D\
  ^1\Pi_\mathrm{u}$} states of hydrogen. {A}b initio calculation of rovibronic
  coupling in {H}$_2$, {HD}, and {D}$_2$},\ }\href
  {https://doi.org/10.1063/1.455271} {\bibfield  {journal} {\bibinfo  {journal}
  {J. Chem. Phys.}\ }\textbf {\bibinfo {volume} {89}},\ \bibinfo {pages} {7401}
  (\bibinfo {year} {1988})}\BibitemShut {NoStop}%
\bibitem [{\citenamefont {Dressler}\ and\ \citenamefont
  {Wolniewicz}(1986)}]{DrWo86}%
  \BibitemOpen
  \bibfield  {author} {\bibinfo {author} {\bibfnamefont {K.}~\bibnamefont
  {Dressler}}\ and\ \bibinfo {author} {\bibfnamefont {L.}~\bibnamefont
  {Wolniewicz}},\ }\bibfield  {title} {\bibinfo {title} {Improved adiabatic
  corrections for the {$B\ ^1\Sigma_\mathrm{u}^+$}, {$C\ ^1\Pi_\mathrm{u}$},
  and $d\ ^1\pi_u$ states of the hydrogen molecule and vibrational structures
  for {H$_2$}, {HD}, and {D$_2$}},\ }\href {https://doi.org/10.1063/1.451040}
  {\bibfield  {journal} {\bibinfo  {journal} {J. Chem. Phys.}\ }\textbf
  {\bibinfo {volume} {85}},\ \bibinfo {pages} {2821} (\bibinfo {year}
  {1986})}\BibitemShut {NoStop}%
\bibitem [{\citenamefont {Quadrelli}\ \emph {et~al.}(1990)\citenamefont
  {Quadrelli}, \citenamefont {Dressler},\ and\ \citenamefont
  {Wolniewicz}}]{QuDrWo90}%
  \BibitemOpen
  \bibfield  {author} {\bibinfo {author} {\bibfnamefont {P.}~\bibnamefont
  {Quadrelli}}, \bibinfo {author} {\bibfnamefont {K.}~\bibnamefont
  {Dressler}},\ and\ \bibinfo {author} {\bibfnamefont {L.}~\bibnamefont
  {Wolniewicz}},\ }\bibfield  {title} {\bibinfo {title} {Nonadiabatic coupling
  between the {EF+GK+H} {$^1\Sigma_\mathrm{g}^+$}, {I} {$^1\Pi_\mathrm{g}$},
  and {J} {$^1\Delta_\mathrm{g}$} states of the hydrogen molecule.
  {C}alculation of rovibronic structures in {H}$_2$, {HD}, and {D}$_2$},\
  }\href {https://doi.org/10.1063/1.458181} {\bibfield  {journal} {\bibinfo
  {journal} {J. Chem. Phys.}\ }\textbf {\bibinfo {volume} {92}},\ \bibinfo
  {pages} {7461} (\bibinfo {year} {1990})}\BibitemShut {NoStop}%
\bibitem [{\citenamefont {Wolniewicz}\ and\ \citenamefont
  {Dressler}(1992)}]{WoDr92}%
  \BibitemOpen
  \bibfield  {author} {\bibinfo {author} {\bibfnamefont {L.}~\bibnamefont
  {Wolniewicz}}\ and\ \bibinfo {author} {\bibfnamefont {K.}~\bibnamefont
  {Dressler}},\ }\bibfield  {title} {\bibinfo {title} {Nonadiabatic energy
  corrections for the vibrational levels of the {$B$} and {$B'\ ^1\Sigma_u^+$}
  states of the {H}$_2$ and {D}$_2$ molecules},\ }\href
  {https://doi.org/10.1063/1.462647} {\bibfield  {journal} {\bibinfo  {journal}
  {J. Chem. Phys.}\ }\textbf {\bibinfo {volume} {96}},\ \bibinfo {pages} {6053}
  (\bibinfo {year} {1992})}\BibitemShut {NoStop}%
\bibitem [{\citenamefont {Wolniewicz}\ and\ \citenamefont
  {Dressler}(1994)}]{WoDr94}%
  \BibitemOpen
  \bibfield  {author} {\bibinfo {author} {\bibfnamefont {L.}~\bibnamefont
  {Wolniewicz}}\ and\ \bibinfo {author} {\bibfnamefont {K.}~\bibnamefont
  {Dressler}},\ }\bibfield  {title} {\bibinfo {title} {Adiabatic potential
  curves and nonadiabatic coupling functions for the first five excited
  {$^1\Sigma_\text{g}^+$} states of the hydrogen molecule},\ }\href
  {https://doi.org/10.1063/1.466957} {\bibfield  {journal} {\bibinfo  {journal}
  {J. Chem. Phys.}\ }\textbf {\bibinfo {volume} {100}},\ \bibinfo {pages} {444}
  (\bibinfo {year} {1994})},\ \Eprint
  {https://arxiv.org/abs/https://doi.org/10.1063/1.466957}
  {https://doi.org/10.1063/1.466957} \BibitemShut {NoStop}%
\bibitem [{\citenamefont {Yu}\ and\ \citenamefont {Dressler}(1994)}]{YuDr94}%
  \BibitemOpen
  \bibfield  {author} {\bibinfo {author} {\bibfnamefont {S.}~\bibnamefont
  {Yu}}\ and\ \bibinfo {author} {\bibfnamefont {K.}~\bibnamefont {Dressler}},\
  }\bibfield  {title} {\bibinfo {title} {Calculation of rovibronic structures
  in the lowest nine excited
  $^1{\Sigma}^+_\text{g}+^1{\Pi}_\text{g}+^1{\Delta}_\text{g}$ states of
  {H}$_2$, {D}$_2$, and {T}$_2$},\ }\href {https://doi.org/10.1063/1.468263}
  {\bibfield  {journal} {\bibinfo  {journal} {J. Chem. Phys.}\ }\textbf
  {\bibinfo {volume} {101}},\ \bibinfo {pages} {7692} (\bibinfo {year}
  {1994})}\BibitemShut {NoStop}%
\bibitem [{\citenamefont {Teufel}(2003)}]{Te03}%
  \BibitemOpen
  \bibfield  {author} {\bibinfo {author} {\bibfnamefont {S.}~\bibnamefont
  {Teufel}},\ }\href@noop {} {\emph {\bibinfo {title} {Adiabatic perturbation
  theory in quantum dynamics}}},\ Lecture Notes in Mathematics\ (\bibinfo
  {publisher} {Springer},\ \bibinfo {year} {2003})\BibitemShut {NoStop}%
\bibitem [{\citenamefont {Panati}\ \emph {et~al.}(2007)\citenamefont {Panati},
  \citenamefont {Spohn},\ and\ \citenamefont {Teufel}}]{PaSpTe07}%
  \BibitemOpen
  \bibfield  {author} {\bibinfo {author} {\bibfnamefont {G.}~\bibnamefont
  {Panati}}, \bibinfo {author} {\bibfnamefont {H.}~\bibnamefont {Spohn}},\ and\
  \bibinfo {author} {\bibfnamefont {S.}~\bibnamefont {Teufel}},\ }\bibfield
  {title} {\bibinfo {title} {The time-dependent born--oppenheimer
  approximation},\ }\href {https://doi.org/10.1051/m2an:2007023} {\bibfield
  {journal} {\bibinfo  {journal} {ESAIM: Mathematical Modelling and Numerical
  Analysis}\ }\textbf {\bibinfo {volume} {41}},\ \bibinfo {pages} {297}
  (\bibinfo {year} {2007})}\BibitemShut {NoStop}%
\bibitem [{\citenamefont {M\'atyus}\ and\ \citenamefont
  {Teufel}(2019)}]{MaTe19}%
  \BibitemOpen
  \bibfield  {author} {\bibinfo {author} {\bibfnamefont {E.}~\bibnamefont
  {M\'atyus}}\ and\ \bibinfo {author} {\bibfnamefont {S.}~\bibnamefont
  {Teufel}},\ }\bibfield  {title} {\bibinfo {title} {Effective non-adiabatic
  {H}amiltonians for the quantum nuclear motion over coupled electronic
  states},\ }\href {https://doi.org/10.1063/1.5097899} {\bibfield  {journal}
  {\bibinfo  {journal} {J. Chem. Phys.}\ }\textbf {\bibinfo {volume} {151}},\
  \bibinfo {pages} {014113} (\bibinfo {year} {2019})}\BibitemShut {NoStop}%
\bibitem [{\citenamefont {Schwartz}(1961)}]{Sc61}%
  \BibitemOpen
  \bibfield  {author} {\bibinfo {author} {\bibfnamefont {C.}~\bibnamefont
  {Schwartz}},\ }\bibfield  {title} {\bibinfo {title} {Lamb shift in the helium
  atom},\ }\href {https://doi.org/10.1103/PhysRev.123.1700} {\bibfield
  {journal} {\bibinfo  {journal} {Phys. Rev.}\ }\textbf {\bibinfo {volume}
  {123}},\ \bibinfo {pages} {1700} (\bibinfo {year} {1961})}\BibitemShut
  {NoStop}%
\bibitem [{\citenamefont {Herman}\ and\ \citenamefont
  {Asgharian}(1966)}]{HeAs66}%
  \BibitemOpen
  \bibfield  {author} {\bibinfo {author} {\bibfnamefont {R.~M.}\ \bibnamefont
  {Herman}}\ and\ \bibinfo {author} {\bibfnamefont {A.}~\bibnamefont
  {Asgharian}},\ }\bibfield  {title} {\bibinfo {title} {Theory of energy shifts
  associated with deviations from born--oppenheimer behavior in
  {$^1\Sigma$}-state diatomic molecules},\ }\href
  {https://doi.org/10.1016/0022-2852(66)90254-2} {\bibfield  {journal}
  {\bibinfo  {journal} {J. Mol. Spectrosc.}\ }\textbf {\bibinfo {volume}
  {19}},\ \bibinfo {pages} {305} (\bibinfo {year} {1966})}\BibitemShut
  {NoStop}%
\bibitem [{\citenamefont {Herman}\ and\ \citenamefont
  {Ogilvie}(1998)}]{HeOg98}%
  \BibitemOpen
  \bibfield  {author} {\bibinfo {author} {\bibfnamefont {R.~M.}\ \bibnamefont
  {Herman}}\ and\ \bibinfo {author} {\bibfnamefont {J.~F.}\ \bibnamefont
  {Ogilvie}},\ }\bibfield  {title} {\bibinfo {title} {An effective
  {H}amiltonian to treat adiabatic and nonadiabatic effects in the rotational
  and vibrational spectra of diatomic molecules},\ }\href
  {https://doi.org/10.1002/9780470141625.ch2} {\bibfield  {journal} {\bibinfo
  {journal} {Adv. Chem. Phys.}\ }\textbf {\bibinfo {volume} {103}},\ \bibinfo
  {pages} {187} (\bibinfo {year} {1998})}\BibitemShut {NoStop}%
\bibitem [{\citenamefont {Pachucki}\ and\ \citenamefont
  {Komasa}(2009)}]{PaKo09}%
  \BibitemOpen
  \bibfield  {author} {\bibinfo {author} {\bibfnamefont {K.}~\bibnamefont
  {Pachucki}}\ and\ \bibinfo {author} {\bibfnamefont {J.}~\bibnamefont
  {Komasa}},\ }\bibfield  {title} {\bibinfo {title} {Nonadiabatic corrections
  to rovibrational levels of {H}$_2$},\ }\href
  {https://doi.org/https://doi.org/10.1063/1.3114680} {\bibfield  {journal}
  {\bibinfo  {journal} {J. Chem. Phys.}\ }\textbf {\bibinfo {volume} {130}},\
  \bibinfo {pages} {164113} (\bibinfo {year} {2009})}\BibitemShut {NoStop}%
\bibitem [{\citenamefont {Scherrer}\ \emph {et~al.}(2017)\citenamefont
  {Scherrer}, \citenamefont {Agostini}, \citenamefont {Sebastiani},
  \citenamefont {Gross},\ and\ \citenamefont {Vuilleumier}}]{prx17}%
  \BibitemOpen
  \bibfield  {author} {\bibinfo {author} {\bibfnamefont {A.}~\bibnamefont
  {Scherrer}}, \bibinfo {author} {\bibfnamefont {F.}~\bibnamefont {Agostini}},
  \bibinfo {author} {\bibfnamefont {D.}~\bibnamefont {Sebastiani}}, \bibinfo
  {author} {\bibfnamefont {E.~K.~U.}\ \bibnamefont {Gross}},\ and\ \bibinfo
  {author} {\bibfnamefont {R.}~\bibnamefont {Vuilleumier}},\ }\bibfield
  {title} {\bibinfo {title} {On the mass of atoms in molecules: Beyond the
  {B}orn--{O}ppenheimer approximation},\ }\href
  {https://doi.org/10.1103/PhysRevX.7.031035} {\bibfield  {journal} {\bibinfo
  {journal} {Phys. Rev. X}\ }\textbf {\bibinfo {volume} {7}},\ \bibinfo {pages}
  {031035} (\bibinfo {year} {2017})}\BibitemShut {NoStop}%
\bibitem [{\citenamefont {Schwenke}(2001{\natexlab{a}})}]{Sch01H2p}%
  \BibitemOpen
  \bibfield  {author} {\bibinfo {author} {\bibfnamefont {D.~W.}\ \bibnamefont
  {Schwenke}},\ }\bibfield  {title} {\bibinfo {title} {A first principle
  effective {H}amiltonian for including nonadiabatic effects for {H}$^+_2$ and
  {HD}$^+$},\ }\href {https://doi.org/10.1063/1.1334897} {\bibfield  {journal}
  {\bibinfo  {journal} {J. Chem. Phys.}\ }\textbf {\bibinfo {volume} {114}},\
  \bibinfo {pages} {1693} (\bibinfo {year} {2001}{\natexlab{a}})}\BibitemShut
  {NoStop}%
\bibitem [{\citenamefont {Schwenke}(2001{\natexlab{b}})}]{Sch01H2O}%
  \BibitemOpen
  \bibfield  {author} {\bibinfo {author} {\bibfnamefont {D.~W.}\ \bibnamefont
  {Schwenke}},\ }\bibfield  {title} {\bibinfo {title} {Beyond the potential
  energy surface: Ab initio corrections to the {B}orn--{O}ppenheimer
  approximation for {H$_2$O}},\ }\href {https://doi.org/10.1021/jp0032513}
  {\bibfield  {journal} {\bibinfo  {journal} {J. Phys. Chem. A}\ }\textbf
  {\bibinfo {volume} {105}},\ \bibinfo {pages} {2352} (\bibinfo {year}
  {2001}{\natexlab{b}})}\BibitemShut {NoStop}%
\bibitem [{\citenamefont {Czachorowski}\ \emph {et~al.}(2018)\citenamefont
  {Czachorowski}, \citenamefont {Puchalski}, \citenamefont {Komasa},\ and\
  \citenamefont {Pachucki}}]{CzPuKoPa18}%
  \BibitemOpen
  \bibfield  {author} {\bibinfo {author} {\bibfnamefont {P.}~\bibnamefont
  {Czachorowski}}, \bibinfo {author} {\bibfnamefont {M.}~\bibnamefont
  {Puchalski}}, \bibinfo {author} {\bibfnamefont {J.}~\bibnamefont {Komasa}},\
  and\ \bibinfo {author} {\bibfnamefont {K.}~\bibnamefont {Pachucki}},\
  }\bibfield  {title} {\bibinfo {title} {Nonadiabatic relativistic correction
  in ${\mathrm{h}}_{2}$, ${\mathrm{d}}_{2}$, and hd},\ }\href
  {https://doi.org/10.1103/PhysRevA.98.052506} {\bibfield  {journal} {\bibinfo
  {journal} {Phys. Rev. A}\ }\textbf {\bibinfo {volume} {98}},\ \bibinfo
  {pages} {052506} (\bibinfo {year} {2018})}\BibitemShut {NoStop}%
\bibitem [{\citenamefont {Komasa}\ \emph {et~al.}(2019)\citenamefont {Komasa},
  \citenamefont {Puchalski}, \citenamefont {Czachorowski}, \citenamefont
  {\L{}ach},\ and\ \citenamefont {Pachucki}}]{KoPuCzLaPa19}%
  \BibitemOpen
  \bibfield  {author} {\bibinfo {author} {\bibfnamefont {J.}~\bibnamefont
  {Komasa}}, \bibinfo {author} {\bibfnamefont {M.}~\bibnamefont {Puchalski}},
  \bibinfo {author} {\bibfnamefont {P.}~\bibnamefont {Czachorowski}}, \bibinfo
  {author} {\bibfnamefont {G.}~\bibnamefont {\L{}ach}},\ and\ \bibinfo {author}
  {\bibfnamefont {K.}~\bibnamefont {Pachucki}},\ }\bibfield  {title} {\bibinfo
  {title} {Rovibrational energy levels of the hydrogen molecule through
  nonadiabatic perturbation theory},\ }\href
  {https://doi.org/10.1103/PhysRevA.100.032519} {\bibfield  {journal} {\bibinfo
   {journal} {Phys. Rev. A}\ }\textbf {\bibinfo {volume} {100}},\ \bibinfo
  {pages} {032519} (\bibinfo {year} {2019})}\BibitemShut {NoStop}%
\bibitem [{\citenamefont {Ferenc}\ and\ \citenamefont
  {Mátyus}(2019)}]{FeMa19HH}%
  \BibitemOpen
  \bibfield  {author} {\bibinfo {author} {\bibfnamefont {D.}~\bibnamefont
  {Ferenc}}\ and\ \bibinfo {author} {\bibfnamefont {E.}~\bibnamefont
  {Mátyus}},\ }\bibfield  {title} {\bibinfo {title} {Non-adiabatic mass
  correction for excited states of molecular hydrogen: Improvement for the
  outer-well ${H}\bar{H}\ ^1{\Sigma_g^+}$ term values},\ }\href
  {https://doi.org/10.1063/1.5109964} {\bibfield  {journal} {\bibinfo
  {journal} {J. Chem. Phys.}\ }\textbf {\bibinfo {volume} {151}},\ \bibinfo
  {pages} {094101} (\bibinfo {year} {2019})}\BibitemShut {NoStop}%
\bibitem [{\citenamefont {Ferenc}\ \emph {et~al.}(2020)\citenamefont {Ferenc},
  \citenamefont {Korobov},\ and\ \citenamefont {M\'atyus}}]{FeKoMa20}%
  \BibitemOpen
  \bibfield  {author} {\bibinfo {author} {\bibfnamefont {D.}~\bibnamefont
  {Ferenc}}, \bibinfo {author} {\bibfnamefont {V.~I.}\ \bibnamefont
  {Korobov}},\ and\ \bibinfo {author} {\bibfnamefont {E.}~\bibnamefont
  {M\'atyus}},\ }\bibfield  {title} {\bibinfo {title} {Nonadiabatic,
  relativistic, and leading-order {QED} corrections for rovibrational intervals
  of ${^{4}\mathrm{He}}_{2}^{+}$ (${X}\text{
  }{^{2}\mathrm{\ensuremath{\Sigma}}}_{u}^{+}$)},\ }\href
  {https://doi.org/10.1103/PhysRevLett.125.213001} {\bibfield  {journal}
  {\bibinfo  {journal} {Phys. Rev. Lett.}\ }\textbf {\bibinfo {volume} {125}},\
  \bibinfo {pages} {213001} (\bibinfo {year} {2020})}\BibitemShut {NoStop}%
\bibitem [{\citenamefont {Ferenc}\ and\ \citenamefont
  {Mátyus}(2022)}]{FeMa22H3}%
  \BibitemOpen
  \bibfield  {author} {\bibinfo {author} {\bibfnamefont {D.}~\bibnamefont
  {Ferenc}}\ and\ \bibinfo {author} {\bibfnamefont {E.}~\bibnamefont
  {Mátyus}},\ }\bibfield  {title} {\bibinfo {title} {Benchmark potential
  energy curve for collinear {H$_3$}},\ }\href@noop {} {\bibfield  {journal}
  {\bibinfo  {journal} {Chem. Phys. Lett.}\ } (\bibinfo {year}
  {2022})}\BibitemShut {NoStop}%
\bibitem [{\citenamefont {Ireland}\ \emph {et~al.}(2021)\citenamefont
  {Ireland}, \citenamefont {Jeszenszki}, \citenamefont {M\'atyus},
  \citenamefont {Martinazzo}, \citenamefont {Ronto},\ and\ \citenamefont
  {Pollak}}]{IrJeMaMaRoPo21}%
  \BibitemOpen
  \bibfield  {author} {\bibinfo {author} {\bibfnamefont {R.}~\bibnamefont
  {Ireland}}, \bibinfo {author} {\bibfnamefont {P.}~\bibnamefont {Jeszenszki}},
  \bibinfo {author} {\bibfnamefont {E.}~\bibnamefont {M\'atyus}}, \bibinfo
  {author} {\bibfnamefont {R.}~\bibnamefont {Martinazzo}}, \bibinfo {author}
  {\bibfnamefont {M.}~\bibnamefont {Ronto}},\ and\ \bibinfo {author}
  {\bibfnamefont {E.}~\bibnamefont {Pollak}},\ }\bibfield  {title} {\bibinfo
  {title} {Lower bounds for atomic energies},\ }\bibfield  {journal} {\bibinfo
  {journal} {ACS Phys. Chem. Au}\ }\href
  {https://doi.org/10.1021/acsphyschemau.1c00018}
  {10.1021/acsphyschemau.1c00018} (\bibinfo {year} {2021})\BibitemShut
  {NoStop}%
\bibitem [{\citenamefont {Mátyus}(2018{\natexlab{a}})}]{Ma18nonad}%
  \BibitemOpen
  \bibfield  {author} {\bibinfo {author} {\bibfnamefont {E.}~\bibnamefont
  {Mátyus}},\ }\bibfield  {title} {\bibinfo {title} {Non-adiabatic mass
  correction to the rovibrational states of molecules. {N}umerical application
  for the {H}$_2^+$ molecular ion},\ }\href {https://doi.org/10.1063/1.5050401}
  {\bibfield  {journal} {\bibinfo  {journal} {J. Chem. Phys.}\ }\textbf
  {\bibinfo {volume} {149}},\ \bibinfo {pages} {194111} (\bibinfo {year}
  {2018}{\natexlab{a}})}\BibitemShut {NoStop}%
\bibitem [{\citenamefont {Mátyus}(2018{\natexlab{b}})}]{Ma18he2p}%
  \BibitemOpen
  \bibfield  {author} {\bibinfo {author} {\bibfnamefont {E.}~\bibnamefont
  {Mátyus}},\ }\bibfield  {title} {\bibinfo {title} {Non-adiabatic
  mass-correction functions and rovibrational states of $^4${H}e$^{+}_2$ ({$X\
  ^2{\Sigma}_\text{u}^+$})},\ }\href {https://doi.org/10.1063/1.5050403}
  {\bibfield  {journal} {\bibinfo  {journal} {J. Chem. Phys.}\ }\textbf
  {\bibinfo {volume} {149}},\ \bibinfo {pages} {194112} (\bibinfo {year}
  {2018}{\natexlab{b}})}\BibitemShut {NoStop}%
\bibitem [{\citenamefont {Mátyus}(2019)}]{Ma19rev}%
  \BibitemOpen
  \bibfield  {author} {\bibinfo {author} {\bibfnamefont {E.}~\bibnamefont
  {Mátyus}},\ }\bibfield  {title} {\bibinfo {title} {Pre-{B}orn--{O}ppenheimer
  molecular structure theory},\ }\href@noop {} {\bibfield  {journal} {\bibinfo
  {journal} {Mol. Phys.}\ }\textbf {\bibinfo {volume} {117}},\ \bibinfo {pages}
  {590} (\bibinfo {year} {2019})}\BibitemShut {NoStop}%
\bibitem [{\citenamefont {Ferenc}\ and\ \citenamefont
  {M\'atyus}(2019)}]{FeMa19EF}%
  \BibitemOpen
  \bibfield  {author} {\bibinfo {author} {\bibfnamefont {D.}~\bibnamefont
  {Ferenc}}\ and\ \bibinfo {author} {\bibfnamefont {E.}~\bibnamefont
  {M\'atyus}},\ }\bibfield  {title} {\bibinfo {title} {Computation of
  rovibronic resonances of molecular hydrogen: ${EF}~^1{\Sigma}_\text{g}^+$
  inner-well rotational states},\ }\href
  {https://doi.org/10.1103/PhysRevA.100.020501} {\bibfield  {journal} {\bibinfo
   {journal} {Phys. Rev. A}\ }\textbf {\bibinfo {volume} {100}},\ \bibinfo
  {pages} {020501(R)} (\bibinfo {year} {2019})}\BibitemShut {NoStop}%
\bibitem [{\citenamefont {M{\'a}tyus}\ and\ \citenamefont
  {{Cassam-Chena{\"i}}}(2021)}]{MaCa21}%
  \BibitemOpen
  \bibfield  {author} {\bibinfo {author} {\bibfnamefont {E.}~\bibnamefont
  {M{\'a}tyus}}\ and\ \bibinfo {author} {\bibfnamefont {P.}~\bibnamefont
  {{Cassam-Chena{\"i}}}},\ }\bibfield  {title} {\bibinfo {title} {Orientational
  decoherence within molecules and emergence of the molecular shape},\ }\href
  {https://doi.org/10.1063/5.0036568} {\bibfield  {journal} {\bibinfo
  {journal} {J. Chem. Phys.}\ }\textbf {\bibinfo {volume} {154}},\ \bibinfo
  {pages} {024114} (\bibinfo {year} {2021})}\BibitemShut {NoStop}%
\bibitem [{\citenamefont {Jeszenszki}\ \emph {et~al.}(2021)\citenamefont
  {Jeszenszki}, \citenamefont {Ireland}, \citenamefont {Ferenc},\ and\
  \citenamefont {M{\'a}tyus}}]{JeIrFeMa21}%
  \BibitemOpen
  \bibfield  {author} {\bibinfo {author} {\bibfnamefont {P.}~\bibnamefont
  {Jeszenszki}}, \bibinfo {author} {\bibfnamefont {R.~T.}\ \bibnamefont
  {Ireland}}, \bibinfo {author} {\bibfnamefont {D.}~\bibnamefont {Ferenc}},\
  and\ \bibinfo {author} {\bibfnamefont {E.}~\bibnamefont {M{\'a}tyus}},\
  }\bibfield  {title} {\bibinfo {title} {On the inclusion of cusp effects in
  expectation values with explicitly correlated {{Gaussians}}},\ }\href
  {https://doi.org/doi.org/10.1002/qua.26819} {\bibfield  {journal} {\bibinfo
  {journal} {Int. J. Quant. Chem.}\ } (\bibinfo {year} {2021})}\BibitemShut
  {NoStop}%
\bibitem [{\citenamefont {Jeszenszki}\ \emph {et~al.}(2022)\citenamefont
  {Jeszenszki}, \citenamefont {Ferenc},\ and\ \citenamefont
  {Mátyus}}]{JeFeMa22}%
  \BibitemOpen
  \bibfield  {author} {\bibinfo {author} {\bibfnamefont {P.}~\bibnamefont
  {Jeszenszki}}, \bibinfo {author} {\bibfnamefont {D.}~\bibnamefont {Ferenc}},\
  and\ \bibinfo {author} {\bibfnamefont {E.}~\bibnamefont {Mátyus}},\
  }\bibfield  {title} {\bibinfo {title} {Variational {D}irac--{C}oulomb
  explicitly correlated computations for atoms and molecules},\ }\href@noop {}
  {\bibfield  {journal} {\bibinfo  {journal} {J. Chem. Phys.}\ } (\bibinfo
  {year} {2022})}\BibitemShut {NoStop}%
\bibitem [{\citenamefont {Ferenc}\ \emph {et~al.}(2022)\citenamefont {Ferenc},
  \citenamefont {Jeszenszki},\ and\ \citenamefont {Mátyus}}]{FeJeMa22}%
  \BibitemOpen
  \bibfield  {author} {\bibinfo {author} {\bibfnamefont {D.}~\bibnamefont
  {Ferenc}}, \bibinfo {author} {\bibfnamefont {P.}~\bibnamefont {Jeszenszki}},\
  and\ \bibinfo {author} {\bibfnamefont {E.}~\bibnamefont {Mátyus}},\
  }\bibfield  {title} {\bibinfo {title} {On the {B}reit interaction in an
  explicitly correlated variational {D}irac--{C}oulomb framework},\ }\href@noop
  {} {\bibfield  {journal} {\bibinfo  {journal} {J. Chem. Phys.}\ } (\bibinfo
  {year} {2022})}\BibitemShut {NoStop}%
\bibitem [{\citenamefont {Cencek}\ and\ \citenamefont
  {Kutzelnigg}(1997)}]{CeKu1997}%
  \BibitemOpen
  \bibfield  {author} {\bibinfo {author} {\bibfnamefont {W.}~\bibnamefont
  {Cencek}}\ and\ \bibinfo {author} {\bibfnamefont {W.}~\bibnamefont
  {Kutzelnigg}},\ }\bibfield  {title} {\bibinfo {title} {Accurate adiabatic
  correction for the hydrogen molecule using the {B}orn--{H}andy formula},\
  }\href {https://doi.org/10.1016/S0009-2614(97)00017-1} {\bibfield  {journal}
  {\bibinfo  {journal} {Chem. Phys. Lett.}\ }\textbf {\bibinfo {volume}
  {266}},\ \bibinfo {pages} {383} (\bibinfo {year} {1997})}\BibitemShut
  {NoStop}%
\bibitem [{\citenamefont {Siłkowski}\ \emph {et~al.}(2021)\citenamefont
  {Siłkowski}, \citenamefont {Zientkiewicz},\ and\ \citenamefont
  {Pachucki}}]{SiZsPa18}%
  \BibitemOpen
  \bibfield  {author} {\bibinfo {author} {\bibfnamefont {M.}~\bibnamefont
  {Siłkowski}}, \bibinfo {author} {\bibfnamefont {M.}~\bibnamefont
  {Zientkiewicz}},\ and\ \bibinfo {author} {\bibfnamefont {K.}~\bibnamefont
  {Pachucki}},\ }\bibfield  {title} {\bibinfo {title} {Chapter {Twelve} -
  {A}ccurate {B}orn--{O}ppenheimer potentials for excited {$\Sigma^+$} states
  of the hydrogen molecule},\ }in\ \href
  {https://doi.org/https://doi.org/10.1016/bs.aiq.2021.05.012} {\emph {\bibinfo
  {booktitle} {New Electron Correlation Methods and their Applications, and Use
  of Atomic Orbitals with Exponential Asymptotes}}},\ \bibinfo {series} {Adv.
  Quant. Chem.}, Vol.~\bibinfo {volume} {83},\ \bibinfo {editor} {edited by\
  \bibinfo {editor} {\bibfnamefont {M.}~\bibnamefont {Musial}}\ and\ \bibinfo
  {editor} {\bibfnamefont {P.~E.}\ \bibnamefont {Hoggan}}}\ (\bibinfo
  {publisher} {Academic Press},\ \bibinfo {year} {2021})\ pp.\ \bibinfo {pages}
  {255--267}\BibitemShut {NoStop}%
\bibitem [{\citenamefont {Reinhold}\ \emph {et~al.}(1999)\citenamefont
  {Reinhold}, \citenamefont {Hogervorst}, \citenamefont {Ubachs},\ and\
  \citenamefont {Wolniewicz}}]{ReHoUbWo99}%
  \BibitemOpen
  \bibfield  {author} {\bibinfo {author} {\bibfnamefont {E.}~\bibnamefont
  {Reinhold}}, \bibinfo {author} {\bibfnamefont {W.}~\bibnamefont
  {Hogervorst}}, \bibinfo {author} {\bibfnamefont {W.}~\bibnamefont {Ubachs}},\
  and\ \bibinfo {author} {\bibfnamefont {L.}~\bibnamefont {Wolniewicz}},\
  }\bibfield  {title} {\bibinfo {title} {Experimental and theoretical
  investigation of the $h\bar{H}\ ^1\sigma_\text{g}^+$ state in {H}$_2$,
  {D}$_2$, and hd, and the $b''\bar{B}\ ^1\sigma_\text{u}^+$ state in {HD}},\
  }\href@noop {} {\bibfield  {journal} {\bibinfo  {journal} {Phys. Rev. A}\
  }\textbf {\bibinfo {volume} {60}},\ \bibinfo {pages} {1258} (\bibinfo {year}
  {1999})}\BibitemShut {NoStop}%
\bibitem [{\citenamefont {Light}\ and\ \citenamefont
  {Carrington~Jr.}(2000)}]{LiCa00}%
  \BibitemOpen
  \bibfield  {author} {\bibinfo {author} {\bibfnamefont {J.~C.}\ \bibnamefont
  {Light}}\ and\ \bibinfo {author} {\bibfnamefont {T.}~\bibnamefont
  {Carrington~Jr.}},\ }\bibinfo {title} {Discrete-variable representations and
  their utilization},\ in\ \href {https://doi.org/10.1002/9780470141731.ch4}
  {\emph {\bibinfo {booktitle} {Advances in Chemical Physics}}}\ (\bibinfo
  {publisher} {John Wiley \& Sons, Ltd},\ \bibinfo {year} {2000})\ pp.\
  \bibinfo {pages} {263--310}\BibitemShut {NoStop}%
\bibitem [{\citenamefont {Pachucki}\ and\ \citenamefont
  {Komasa}(2018)}]{PaKo18}%
  \BibitemOpen
  \bibfield  {author} {\bibinfo {author} {\bibfnamefont {K.}~\bibnamefont
  {Pachucki}}\ and\ \bibinfo {author} {\bibfnamefont {J.}~\bibnamefont
  {Komasa}},\ }\bibfield  {title} {\bibinfo {title} {Nonadiabatic rotational
  states of the hydrogen molecule},\ }\href
  {https://doi.org/10.1039/c7cp06516g} {\bibfield  {journal} {\bibinfo
  {journal} {Phys. Chem. Chem. Phys.}\ }\textbf {\bibinfo {volume} {20}},\
  \bibinfo {pages} {247} (\bibinfo {year} {2018})}\BibitemShut {NoStop}%
\bibitem [{\citenamefont {Bailly}\ \emph {et~al.}(2010)\citenamefont {Bailly},
  \citenamefont {Salumbides}, \citenamefont {Vervloet},\ and\ \citenamefont
  {Ubachs}}]{BaSaVeUb10}%
  \BibitemOpen
  \bibfield  {author} {\bibinfo {author} {\bibfnamefont {D.}~\bibnamefont
  {Bailly}}, \bibinfo {author} {\bibfnamefont {E.~J.}\ \bibnamefont
  {Salumbides}}, \bibinfo {author} {\bibfnamefont {M.}~\bibnamefont
  {Vervloet}},\ and\ \bibinfo {author} {\bibfnamefont {W.}~\bibnamefont
  {Ubachs}},\ }\bibfield  {title} {\bibinfo {title} {Accurate level energies in
  the {$EF\ ^1\Sigma_\text{g}^+$}, {$GK\ ^1\Sigma_\text{g}^+$}, {$H\
  ^1\Sigma_\text{g}^+$}, {$B\ ^1\Sigma_\text{u}^+$}, {$C\ ^1\Pi_\text{u}$},
  {$B'\ ^1\Sigma_\text{u}^+$}, {$D\ ^1\Pi_\text{u}$}, {$I\ ^1\Pi_\text{g}$},
  {$J\ ^1\Delta_\text{g}$} states of {H}$_2$},\ }\href
  {https://doi.org/10.1080/00268970903413350} {\bibfield  {journal} {\bibinfo
  {journal} {Mol. Phys.}\ }\textbf {\bibinfo {volume} {108}},\ \bibinfo {pages}
  {827} (\bibinfo {year} {2010})}\BibitemShut {NoStop}%
\bibitem [{\citenamefont {Dickenson}\ \emph {et~al.}(2012)\citenamefont
  {Dickenson}, \citenamefont {Salumbides}, \citenamefont {Niu}, \citenamefont
  {Jungen}, \citenamefont {Ross},\ and\ \citenamefont
  {Ubachs}}]{DiSaNiJuRoUb12}%
  \BibitemOpen
  \bibfield  {author} {\bibinfo {author} {\bibfnamefont {G.~D.}\ \bibnamefont
  {Dickenson}}, \bibinfo {author} {\bibfnamefont {E.~J.}\ \bibnamefont
  {Salumbides}}, \bibinfo {author} {\bibfnamefont {M.}~\bibnamefont {Niu}},
  \bibinfo {author} {\bibfnamefont {C.}~\bibnamefont {Jungen}}, \bibinfo
  {author} {\bibfnamefont {S.~C.}\ \bibnamefont {Ross}},\ and\ \bibinfo
  {author} {\bibfnamefont {W.}~\bibnamefont {Ubachs}},\ }\bibfield  {title}
  {\bibinfo {title} {Precision spectroscopy of high rotational states in
  {H}$_2$ investigated by {D}oppler-free two-photon laser spectroscopy in the
  {$EF\ ^1\Sigma_\text{g}^+$}--{$X\ ^1\Sigma_\text{g}^+$} system},\ }\href
  {https://doi.org/10.1103/PhysRevA.86.032502} {\bibfield  {journal} {\bibinfo
  {journal} {Phys. Rev. A}\ }\textbf {\bibinfo {volume} {86}},\ \bibinfo
  {pages} {032502} (\bibinfo {year} {2012})}\BibitemShut {NoStop}%
\bibitem [{\citenamefont {Hölsch}\ \emph {et~al.}(2018)\citenamefont
  {Hölsch}, \citenamefont {Beyer},\ and\ \citenamefont {Merkt}}]{HoBeMe18}%
  \BibitemOpen
  \bibfield  {author} {\bibinfo {author} {\bibfnamefont {N.}~\bibnamefont
  {Hölsch}}, \bibinfo {author} {\bibfnamefont {M.}~\bibnamefont {Beyer}},\
  and\ \bibinfo {author} {\bibfnamefont {F.}~\bibnamefont {Merkt}},\ }\bibfield
   {title} {\bibinfo {title} {Nonadiabatic effects on the positions and
  lifetimes of the low-lying rovibrational levels of the {GK}
  {$^1\Sigma_\mathrm{g}^+$} and {H} {$^1\Sigma_\mathrm{g}^+$} states of
  {H}$_2$},\ }\href {https://doi.org/10.1039/C8CP05233F} {\bibfield  {journal}
  {\bibinfo  {journal} {Phys. Chem. Chem. Phys.}\ }\textbf {\bibinfo {volume}
  {20}},\ \bibinfo {pages} {26837} (\bibinfo {year} {2018})}\BibitemShut
  {NoStop}%
\bibitem [{\citenamefont {Ferenc}\ and\ \citenamefont
  {Mátyus}()}]{FeMa22bethe}%
  \BibitemOpen
  \bibfield  {author} {\bibinfo {author} {\bibfnamefont {D.}~\bibnamefont
  {Ferenc}}\ and\ \bibinfo {author} {\bibfnamefont {E.}~\bibnamefont
  {Mátyus}},\ }\bibfield  {title} {\bibinfo {title} {Computation of the
  {B}ethe logarithm for polyatomic and polyelectronic molecular systems (in
  preparation)},\ }\href@noop {} {\ }\BibitemShut {NoStop}%
\bibitem [{\citenamefont {Corongiu}\ and\ \citenamefont
  {Clementi}(2009)}]{CoCl09}%
  \BibitemOpen
  \bibfield  {author} {\bibinfo {author} {\bibfnamefont {G.}~\bibnamefont
  {Corongiu}}\ and\ \bibinfo {author} {\bibfnamefont {E.}~\bibnamefont
  {Clementi}},\ }\bibfield  {title} {\bibinfo {title} {Energy and density
  analyses of the {H}$_2$ molecule from the united atom to dissociation: {T}he
  {$^1\Sigma_\text{g}^+$} states},\ }\href {https://doi.org/10.1063/1.3168506}
  {\bibfield  {journal} {\bibinfo  {journal} {J. Chem. Phys.}\ }\textbf
  {\bibinfo {volume} {131}},\ \bibinfo {pages} {034301} (\bibinfo {year}
  {2009})}\BibitemShut {NoStop}%
\bibitem [{\citenamefont {Corongiu}\ and\ \citenamefont
  {Clementi}(2010)}]{CoCl10}%
  \BibitemOpen
  \bibfield  {author} {\bibinfo {author} {\bibfnamefont {G.}~\bibnamefont
  {Corongiu}}\ and\ \bibinfo {author} {\bibfnamefont {E.}~\bibnamefont
  {Clementi}},\ }\bibfield  {title} {\bibinfo {title} {Energy and density
  analysis on the {H}$_2$ molecule from the united atom to dissociation: {T}he
  {$\Sigma$}, {$\Pi$}, {$\Delta$}, {$\Phi$}, and {$\Gamma$} manifolds},\ }\href
  {https://doi.org/10.1002/qua.22831} {\bibfield  {journal} {\bibinfo
  {journal} {Int. J. Quant. Chem.}\ }\textbf {\bibinfo {volume} {111}},\
  \bibinfo {pages} {3517} (\bibinfo {year} {2010})}\BibitemShut {NoStop}%
\bibitem [{\citenamefont {Beyer}\ and\ \citenamefont {Merkt}(2016)}]{BeMe16b}%
  \BibitemOpen
  \bibfield  {author} {\bibinfo {author} {\bibfnamefont {M.}~\bibnamefont
  {Beyer}}\ and\ \bibinfo {author} {\bibfnamefont {F.}~\bibnamefont {Merkt}},\
  }\bibfield  {title} {\bibinfo {title} {Structure and dynamics of {H}$_2^+$
  near the dissociation threshold: A combined experimental and computational
  investigation},\ }\href {https://doi.org/10.1016/j.jms.2016.08.001}
  {\bibfield  {journal} {\bibinfo  {journal} {J. Mol. Spectrosc.}\ }\textbf
  {\bibinfo {volume} {330}},\ \bibinfo {pages} {147} (\bibinfo {year}
  {2016})}\BibitemShut {NoStop}%
\end{thebibliography}
%

\end{document}